\documentclass[english]{article}
\usepackage[T1]{fontenc}
\usepackage[latin9]{inputenc}
\usepackage{geometry}
\usepackage{textcomp}
\usepackage{amstext}
\usepackage{graphicx}

\makeatletter
\newenvironment{lyxcode}
{\par\begin{list}{}{
\setlength{\rightmargin}{\leftmargin}
\setlength{\listparindent}{0pt}% needed for AMS classes
\raggedright
\setlength{\itemsep}{0pt}
\setlength{\parsep}{0pt}
\normalfont\ttfamily}%
 \item[]}
{\end{list}}

\makeatother

\usepackage{babel}
\begin{document}

\title{Practical Magick with C, PDL and PDL::PP -- a guide to compiled add-ons
for PDL}

\author{Craig DeForest, Karl Glazebrook}

\maketitle

\section*{Preface}

This guide is intended to knit together, and extend, the existing
PP and C documentation on PDL internals. It draws heavily from prior
work by the authors of the code. Special thanks go to Christian Soeller,
and Tuomas Lukka, who together with Glazebrook conceived and implemented
PDL and PP; and to Chris Marshall, who has led the PDL development
team through several groundbreaking releases and to new levels of
usability.

\pagebreak{}

\tableofcontents{}

\pagebreak{}

\section{\label{sec:Introduction}Introduction}

PDL uses C to make ``the magic'' happen. PDL variables are represented
internally with C structures that carry metadata and the actual data
around. You can access those directly with C code that you link to
Perl using the perlXS mechanism that is supplied with Perl. 

Handling PDL variables directly in C can be tedious, because it involves
variable-depth nested looping and direct (and fiddly) pointer access
to variable-type arrays. So most PDL functions are written in a metalanguage
called PP that wraps your C code snippet with a concise description
of how a PDL operator should work. PP is compiled by a macro preprocessor,
contained in the module PDL::PP. PDL::PP generates Perl and perlXS
code from your concise description; that code is then compiled and
dynamically linked using perlXS. We use the word \emph{operator} to
distinguish PP-compiled code from a Perl (or XS) \emph{function}.
A PP-compiled operator can include several ancillary Perl and/or C
functions. 

PDL and related vectorized languages like IDL or NumPy work fast because
they automatically vectorize large operations out of the high level
interpreter: while the language itself is processed by an intepreter,
the ``hot spot'' operations at the core of a scientific calculation
run in faster compiled C loops. But all of them are generally slower
than custom written C code. The reason for that is all vectorized
languages access memory in \emph{pessimal} order. If you write an
expression like ``\texttt{\small{}\$a = sqrt( \$b{*}\$b + \$c{*}\$c
)}'', for example, then all of the elements of \$c are accessed twice
and written to a temporary variable; then likewise for \$b; then all
elements of both temporary variables are accessed and the result is
written to yet another temporary variable, which is accessed and processed
with the \texttt{\small{}sqrt} operation -- making 11 passes through
your data. If the data are large enough, then every single operation
will break your CPU's cache and require (slow) access to the system
RAM. The same operation, written in PP, can carry out the root-sum-squares
on each element in order, preserving CPU cache and running \emph{several}
\emph{times} faster.

The point of PP is to make writing custom operators in C \emph{inside}
the vectorized threading loops nearly as quick and easy as writing
Perl expressions. PP handles most of the argument processing and looping
logic that is required to make PDL operations work, so that you only
write the interior of the loop. PP code is written in C with some
extra macros to access the data and metadata of the arguments. With
a small amount of effort, you can dash off new operators almost as
fast as you could write them in PDL itself. The point of this document
is to give you the ability to do that.

This document is intended to be a comprehensive introduction and reference
for PDL data structures and the PP language, unifying and extending
much of the half-written documentation that already exists (late 2013).
Since PP is basically C, you should be familiar with Perl, PDL, and
C. You should also have at least read the \emph{perlguts} man page
and preferably be familiar with XS (\emph{perlxs}), which is Perl's
interface to C.

\subsection{\label{sub:Review-of-PDL}Review of PDL Dimensions}

PDL (the language) manipulates structured-array objects. Each PDL/piddle
(object) has a \emph{dimlist} that lists 0 or more dimensions and
a size for each one. If the dimlist has 0 elements, the piddle is
a ``scalar piddle''. Each element of the dimlist is the size of
the corresponding dimension of the array, and can be 0 or more. If
any element of the dimlist contains 0, then the piddle is \emph{empty}
and contains zero elements. 

There is a special case -- a \emph{null piddle} -- that contains no
dimlist at all. Null piddles are placeholders, used mainly for output
in PP functions. When PP code encounters a null piddle it automatically
resizes and allocates the piddle to the appropriate size for the current
operation, based on the shapes of the other piddles in the same function
call.

Each PDL function or operator operates on objects with some dimension
-- 0, 1, 2, etc. These dimensions are taken from the \emph{front}
of the operands' dimlists, and are called \emph{active dimensions
}or \emph{active dims }of the operator to distinguish them from \emph{thread
dimensions} of a particular operation (Thread dimensions are looped
over automagically by the PP engine, which is what makes PDL a \emph{vector}
language). Scalar operators like + have 0 active dimensions. Matrix
multiplication has 2 active dimensions. Operators with more than one
argument can have different numbers of active dimensions for each
argument -- for example, \texttt{\small{}index} has one active dim
in the source variable (since you must have a dimension to index over)
and zero active dims in the index variable (since the value used to
do the indexing is itself a scalar). Further, the output can have
a different number of active dims than the input. The \emph{collapse
operators}, like \texttt{\small{}sumover} and \texttt{\small{}average},
have one active dim on their input and zero active dims on their output.

Properly written PDL operators access their arguments as if they only
had the active dimensions. A scalar operator treats its arguments
as if they were scalars. A 1-D operator treats all of its arguments
as if they had only a single dimension, etc. Higher dimensions, if
any, are called \emph{thread dims} because they are matched up and
handled by the threading engine.

\subsection{\label{sub:Threading-Rules}Threading Rules}

Threading is a form of automatic vectorization. After the active dimensions
in each argument to an operator are accounted for, any remaining dimensions
are looped over by the PP engine. These dims are called \emph{thread
dims}. All the thread dims in a given operator call must match. They
are handled with three simple rules:
\begin{enumerate}
\item \emph{Missing dimensions are added, with size 1}. This deals with
a basic ambiguity between an array of length 1 and a scalar, by implictly
extending low-dimensional arrays out into higher dimensions with size
1.
\item \emph{Dimensions of size 1 are silently repeated.} This extends the
concept of scalar operators in a vector space (e.g. scalar multiplication,
scalar addition).
\item \emph{All dimensions that are not size 1 must match sizes exactly}.
This prevents trying to add, say, a 2-vector to a 3-vector, which
doesn't make sense. Note that $0\neq1$, so it is possible for an
empty piddle to fail to mix with non-empty piddles.
\end{enumerate}
The threading rules are implemented by code generated by the PP suite
for each PP function/operator. The PP code ``knows'' the difference
between active and thread dimensions by inspecting the function's
\emph{signature} (§\ref{sub:The-signature-of}), which is a compact
way of expressing the structure and dimensionality of an operator. 

There is an additional rule of threading: under normal circumstances
there should be \emph{no causal link }between different iterations
of the threadloop -- i.e. the different computations carried out by
the thread loops must be completely independent of one another. This
leaves the threading engine open to use multiple CPU cores for a large
thread operation. (If you must break this rule for a particular operator,
it is possible to mark the operator as non-parallelizable.)

\subsection{\label{sub:What-is-PDL::PP?}What is PDL::PP?}

PDL::PP is a collection of codewalker/compiler modules that compile
PP definitions into XS and Perl code. The XS code normally then gets
treated with \emph{perlxs} and turned into C code, which in turn gets
compiled and linked into a running Perl. 

Because PP is a \emph{compiler} and not a runtime interpreter, the
compiled code runs quickly. But it also has certain limitations that
aren't present in an interpreter. For example, PP mocks up dynamic
typing. It keeps track of a single data type of the entire calculation,
and autopromotes all the arguments to the type of the strongest one
(except as you specify, see §\ref{sub:Signature-data-types}). Because
the dynamic data types are resolved into a single type before your
code gets called, PP can simply compile multiple copies of your code,
one for each possible data type. When the Perl side user calls your
PP operator, all the arguments get regularized and the appropriately-typed
version of your compiled code gets used. Each PP operator gets compiled
into a \emph{family} of C functions that implement it.

Like Perl code, PP operators can be written into a standalone module,
or placed inline in a Perl script with the \texttt{\small{}Inline::PdlPP}
module. 

PDL::PP is necessary to \emph{build} PP modules, but not to \emph{run}
them -- though this distinction is blurred if you use \texttt{\small{}Inline::PdlPP}
to insert PP code directly in the middle of a Perl script.

\subsection{\label{sub:A-trivial-example}A trivial example: a linear function}

Here's a simple PP routine that performs linear scaling: given input
variables \emph{a}, \emph{b}, and \emph{c}, it returns the value \emph{ab+c}.
This function is nearly universally useful: with appropriate values
of \emph{b} and \emph{c} it can convert between Fahrenheit and Celsius,
scale between samples of different population size, or (if \emph{b}
is a gain matrix and \emph{c} is a dark image value) flat-field a
scientific image. The following is a snippet of \emph{Perl} code,
suitable for pasting directly into a Perl/PDL script or autoload file.
It invokes the module \texttt{\footnotesize{}Inline::Pdlpp} to insert
code in a different language (PP) directly into your script. The PP
code is automagically compiled and linked into your running Perl.
\begin{lyxcode}
{\small{}no~PDL::NiceSlice;~~\#~prevents~interference}{\small \par}

{\small{}use~Inline~Pdlpp~=>~<\textcompwordmark{}<'EOPP';}{\small \par}

{\small{}pp\_def('linscale',}{\small \par}
\begin{lyxcode}
{\small{}Pars~=>~'a();~b();~c();~{[}o{]}o()',~//~Pars~(signature)~specs~threadable~arguments.}{\small \par}

{\small{}Code~=><\textcompwordmark{}<'EOC',}{\small \par}

{\small{}\$o()~=~\$a()~{*}~\$b()~+~\$c();}{\small \par}
\end{lyxcode}
{\small{}EOC}{\small \par}

{\small{});}{\small \par}

{\small{}EOPP}{\small \par}

{\small{}{*}linscale~=~\textbackslash{}\&PDL::linscale;~\#~copy~the~defined~code~into~the~current~package}{\small \par}
\end{lyxcode}
The Inline code declares a PP operator called ``linscale'' that
is defined in the PDL package. definition takes the form of a call
to a function called \texttt{\small{}pp\_def}, which assembles an
operator called \texttt{\small{}PDL::linscale}. The \texttt{\small{}pp\_def}
constructor accepts an operator name, followed by a hash containing
declarative fields. The minimum fields required are the \texttt{\small{}Pars}
section, which contains a concise description of the parameters, and
the \texttt{\small{}Code} section, which contains the actual code
to build. 

This \texttt{\small{}Pars} section specifies three input parameters
with zero active dims each; and one output parameter. The user can
pass in either three PDLs or four. If the fourth (marked \texttt{\small{}{[}o{]}}
for output) is omitted, then the operator will autocreate it and return
it. The \texttt{\small{}Code} section contains a small snippet of
C code with macro substitutions, that carries out the operation for
a single iteration of any thread loops. In this case the C code snippet
is very simple indeed: it just carries out the linear operation and
stuffs the result into the output variable.

You can call this function just by invoking it, as in:
\begin{lyxcode}
{\small{}print~(~~linscale(pdl(1,~2,~3),~2,~pdl(4,~5,~6))~~).''\textbackslash{}n'';}{\small \par}
\end{lyxcode}
which is equivalent to
\begin{lyxcode}
print~(~~pdl(1,2,3){*}2~+~pdl(4,5,6)~~).''\textbackslash{}n'';
\end{lyxcode}
in normal PDL. Either one will generate the output \texttt{\small{}{[}6
9 12{]}}. The PP version of even this trivial example runs slightly
faster, because the high level expression generates several temporary
variables while the PP code walks through the input parameters exactly
once.

\subsection{\label{sub:An-early-example:}A non-trivial example: Mandelbrot set
visualization}

Here's a fast example PP routine for calculating something useful:
whether a particular complex number is in the famous Mandelbrot set,
or no. The Mandelbrot operator on a complex number \emph{z} is given
by $M_{i}(z)=M_{i-1}(z)^{2}+z$, with $M_{0}(z)=z$. The Mandelbrot
set is the set of all \emph{z} such that $|M_{\infty}(z)|$ is bounded.
Visualizing the Mandelbrot set requires testing each pixel for membership
in the set, by iterating $M$ to some large \emph{i}. You could make
a visualizer with simple threading of elementary operators like this:
\begin{lyxcode}
{\small{}sub~thread\_mandel~\{~}{\small \par}
\begin{lyxcode}
{\small{}my~\$z~=~shift;~my~\$cxy~=~\$z->copy;~\#~expects~N=2~in~dim~0~(real,imaginary)}{\small \par}

{\small{}my~\$n~=~shift~//~1000;~~~~~~~~~~~~~\#~max~iteration~count~-{}-~defaults~to~1000.}{\small \par}

{\small{}for~my~\$i(0..\$n)~\{}{\small \par}
\begin{lyxcode}
{\small{}my~\$tmp~=~\$cxy~{*}~\$cxy;}{\small \par}

{\small{}\$cxy->((1))~{*}=~2~{*}~\$cxy->((0));~~~~~~~~~~~~\#~imaginary~part}{\small \par}

{\small{}\$cxy->((0))~.=~\$tmp->((0))~-~\$tmp->((1));~~\#~real~part}{\small \par}

{\small{}\$cxy~+=~\$z;}{\small \par}
\end{lyxcode}
{\small{}\}}{\small \par}

{\small{}return~(((\$cxy{*}\$cxy)->sumover~-~4)->clip(0)->log);}{\small \par}
\end{lyxcode}
{\small{}\}}{\small \par}
\end{lyxcode}
You feed a 2$\times$W$\times$H array of complex numbers to \texttt{\small{}thread\_mandel},
and it returns a W$\times$H PDL indicating whether each input number
\emph{might }be in the Mandelbrot set. If the return value is 0, the
corresponding input number might be in the set; if it's positive,
the corresponding number is out of the set, and the value is a measure
of its distance from the set. The routine uses the threading engine
to operate on an entire complex ``image''. This threaded example
works much faster than a Perl script, but much slower than it could. 

Ideally, we'd like the threading engine to do all the operations for
each point at once\emph{ }(and \emph{then} move on to the next point),
to preserve CPU cache. Also, we'd like it to stop with each point
as soon as it can, to save time on points that diverge fast. Using
\texttt{\footnotesize{}Inline::Pdlpp}, we can do just that -- from
within our Perl script! Right under the \texttt{\footnotesize{}thread\_mandel}
declaration (or anywhere else in our script or module), we can write
this (notice that we ``\texttt{\footnotesize{}use Inline}'' with
arguments, rather than calling \texttt{\footnotesize{}Inline::Pdlpp}
directly):
\begin{lyxcode}
{\small{}no~PDL::NiceSlice;~~\#~prevents~interference}{\small \par}

{\small{}use~Inline~Pdlpp~=>~<\textcompwordmark{}<'EOPP';}{\small \par}

{\small{}pp\_def('pp\_mandel',}{\small \par}
\begin{lyxcode}
{\small{}Pars~=>~'c(n=2);~{[}o{]}o()',~~~~~//~Pars~(signature)~specs~threadable~arguments.}{\small \par}

{\small{}OtherPars~=>~'int~max\_it',~~~~//~OtherPars~specs~scalar~arguments.}{\small \par}

{\small{}Code~=><\textcompwordmark{}<'EOC',}{\small \par}

{\small{}/{*}~All~this~code~gets~wrapped~automagically~in~a~thread~loop.~~~~{*}/}{\small \par}

{\small{}/{*}~It~starts~a~fresh~C~block,~so~you~can~declare~stuff~up~top.~~~{*}/}{\small \par}

{\small{}/{*}~The~\$GENERIC()~macro~is~the~overall~expression~type.~~~~~~~~~~{*}/}{\small \par}

{\small{}int~i;~~~~~~~~~~~~~~~~~~~~~~~~~~~~~~~~~~~~~~~~~~//~iterator.~~}{\small \par}

{\small{}\$GENERIC()~~~rp0~=~\$c(n=>0),~~~ip0~=~\$c(n=>1);~~//~Copy~the~initial~value~to~rp0/ip0}{\small \par}

{\small{}\$GENERIC()~~~rp~~=~rp0,~~~~~~~~ip~~=~ip0;~~~~~~~//~Copy~again~for~the~initial~iteration}{\small \par}

{\small{}\$GENERIC()~~~rp2~=~rp{*}rp,~~~~~~ip2~=~ip{*}ip;~~~~~//~find~RP\textasciicircum{}2~and~IP\textasciicircum{}2~for~magnitude~and~z\textasciicircum{}2.}{\small \par}

{\small{}for(i=\$COMP(max\_it);~rp2+ip2~<~4~\&\&~i;~i-{}-)~\{~~~//~the~OtherPars~are~in~the~\$COMP~macro.}{\small \par}
\begin{lyxcode}
{\small{}ip~{*}=~2~{*}~rp;~~rp~~=~rp2~-~ip2;~~//~calculate~M\_i(z)\textasciicircum{}2}{\small \par}

{\small{}rp~+=~rp0;~~~~~ip~+=~ip0;~~~~~~~~//~add~z}{\small \par}

{\small{}rp2~=~rp{*}rp;~~~ip2~=~ip{*}ip;~~~~~~//~calculate~rp\textasciicircum{}2~and~ip\textasciicircum{}2~for~next~time}{\small \par}
\end{lyxcode}
{\small{}\}}{\small \par}

{\small{}\$o()=~i;~~//~Assign~the~iterator~to~the~output~value}{\small \par}
\end{lyxcode}
{\small{}EOC}{\small \par}

{\small{});}{\small \par}

EOPP

\end{lyxcode}
That code snippet (which can be placed anywhere you'd place a regular
Perl subroutine declaration) declares a function called \texttt{\small{}PDL::pp\_mandel}.
It gets compiled into C and then into object code, and linked into
the running Perl. The C-like code in the core of the declaration \emph{is},
in fact, C -- with some preprocessor macros. It handles one instance
of the function. It gets wrapped automatically with all the required
interface logic and loops to become part of a threaded operator and
interface to perl, as a function in the \texttt{\small{}PDL} package.
\texttt{\small{}PDL::pp\_mandel} accepts two inputs: a PDL (called
'\texttt{\small{}c}' in the code) with a single \emph{active dim}
of size 2; and a separate scalar integer ('\texttt{\small{}max\_it}')
saying how many times (at most) to iterate $M_{i}$ for each point.
The PDL arguments (both input and output) are described in the ``\texttt{\small{}Pars}''
section of the declaration, which also points out that the return
value is a scalar -- the active dim is collapsed and discarded, so
a 2$\times$\emph{W}$\times$\emph{H} input PDL will yield a \emph{W}$\times$\emph{H}
output PDL. Since \texttt{\small{}max\_it} doesn't participate in
the threading operation, it is declared in the ``\texttt{\small{}OtherPars}''
section and gets accessed slightly differently from the main threading
variables. 

The C code is passed into the declaration as a string, and it has
three important macros that are used here:
\begin{itemize}
\item \texttt{\small{}\$GENERIC()} is a stand-in for whatever the type of
the arguments may be. PP compiles a separate variant of the code for
each possible data type of its PDL arguments, so that (e.g.) double
type variables run in a version of the code where \texttt{\small{}\$GENERIC()}
is ``\texttt{\small{}double}''; but byte PDLs run in a version where
\texttt{\small{}\$GENERIC()} is ``\texttt{\small{}unsigned char}''.
\item \texttt{\small{}\$COMP(max\_it)} is an expression that retrieves the
integer parameter. The \texttt{\small{}\$COMP()} macro is used to
store a bunch of compiled-in values specific to each PP function -
it's described more later. All the \texttt{\small{}OtherPars} (which
are typically simple C types) can be found here.
\item \texttt{\small{}\$c()} and \texttt{\small{}\$o()} give access to the
PDL variables declared in the \texttt{\small{}Pars} section. Each
active dim is declared up in the Pars section, and you can specify
which element of each active dim you want, by naming it (as in ``\texttt{\small{}n=>0}'').
There's another construct, called \texttt{\small{}loop}, that you
can use to loop over a particular active dim. Inside a \texttt{\small{}loop},
you don't have to access a particular location along each active dim
-- it's done for you automagically. Here we're not explicitly looping
over any dimension -- the only active dim runs across (real, imaginary),
and we access the components with the \texttt{\small{}``n=>}'' nomenclature.
The parameter name (``\texttt{\small{}n}'' here) is \emph{required,}
in order to avoid confusion in positional notation: sometimes there
is more than one active dim, and/or some of the active dims might
be masked by a surrounding \texttt{\small{}loop} structure.
\end{itemize}
After \texttt{\small{}mandel} is declared (either inline or as part
of a module declaration; see below for details) you can call it just
like any other PDL method -- for example,
\begin{lyxcode}
{\small{}\$vals~=~(ndcoords(200,200)/50-2)->pp\_mandel(1000);}{\small \par}
\end{lyxcode}
will feed a \texttt{\small{}2x200x200} array into \texttt{\small{}PDL::pp\_mandel}
and return a 200$\times$200 array in \texttt{\small{}\$vals}. On
return, any point that \emph{might} be in the Mandelbrot set gets
a return value of 0, and any point that is definitely not in the set
gets the number of remaining iterations at the time it diverged. Unlike
the simple threading version, \texttt{\small{}pp\_mandel} doesn't
waste iterations on numbers that have already diverged.

You can visualize a piece of the mandelbrot set by enumerating a region
on the complex plane and feeding it to \texttt{\small{}pp\_mandel},
then plotting the result:
\begin{lyxcode}
{\small{}pdl>~\$cen~=~pdl(-0.74897,0.05708);}{\small \par}

{\small{}pdl>~\$coords~=~\$cen~+~(ndcoords(501,501)/250~-~1)~{*}~0.001;}{\small \par}

{\small{}pdl>~use~PDL::Graphics::Simple;~}{\small \par}

{\small{}pdl>~\$w=pgswin(size=>{[}600,600{]});}{\small \par}

{\small{}pdl>~\$w->image(~\$coords->using(0,1),~\$coords->pp\_mandel(2500),~\{title=>''Mandelbrot''\}~);}{\small \par}
\end{lyxcode}
yields the image:

\includegraphics[width=4in]{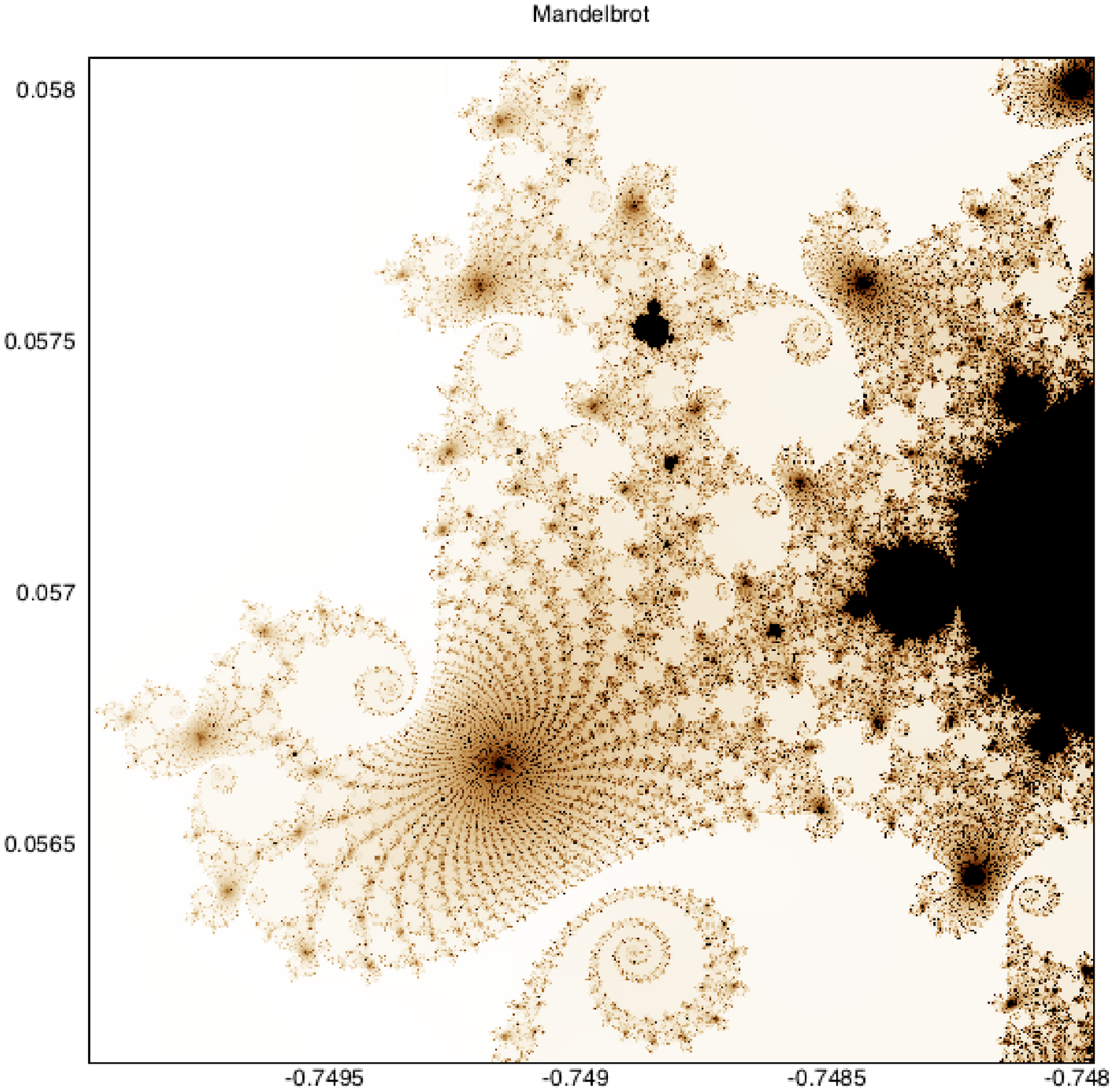}

\section{\label{sec:Basic-PP}Basic PP}

The PP compiler is implemented as a Perl module (\texttt{\small{}PDL::PP}).
A PP source code file is itself a Perl script that loads the \texttt{\small{}PDL::PP}
module and executes methods in it, to generate .pm and .xs output
files. The script calls particular functions exported by the \texttt{\small{}PDL::PP}
module to generate segments of code. Each function call takes some
parameters that describe the particular code segment. The core of
the system is the function \texttt{\small{}pp\_def()}, which defines
a PP operator; but there are many definition functions that let you
mix and match Perl code segments (and POD documentation), PP operators,
and PerlXS / C code segments. 

If you want to produce a standalone module, you put the PP calls into
a Perl script (standard suffix ``.pd'') inside your usual module
directory structure. The script can use PDL::PP itself, but more typically
that is handled by MakeMaker and/or your makefile (see Section \ref{sub:A-standalone-PP}
for a description of how to set up a PP module with MakeMaker). 

You don't have to make a standalone module. The module Inline::PP
can insert some compiled inline PP code directly into your Perl script.
The \texttt{\small{}Inline::Pdlpp} module takes care of using \texttt{\small{}PDL::PP}
and generating the intermediate files for linking, so you can just
insert some pp\_def calls in a ``use Inline Pdlpp'' segment. That
is covered in Section \ref{sub:An-Inline::Pdlpp-Call}.

PP operators do not return values in the conventional way. In general,
you declare one or more named PDL parameters that are flagged as output
variables, and these get autogenerated by PP (See the code in §\ref{sub:A-trivial-example}
and §\ref{sub:An-early-example:} for examples of how this looks).
Your code assigns to particular values inside the output PDL. If you
need to return a non-PDL value, the easiest way is to declare a Perl
helper function that passes in a Perl SV and or SV ref, and modify
that in-place. Then the Perl helper can return the non-PDL value in
the usual way.

\subsection{\label{sub:The-signature-of}Dimensionality and the signature of
an operator}

PP keeps track of the dimensionality of each operator with a ``signature''.
PP operator signatures are similar to, but slightly different from,
C argument declarations. The signature is a list of specifications
and names for PDL arguments to the operator. The arguments are delimited
with ';'. Each specifier is an optional type specifier, followed by
a comma-delimited set of option flags in square brackets, followed
by a variable name and a comma-delimited collection of zero or more
active dim symbols, all enclosed in parens '(',')'. An example:
\begin{lyxcode}
{\small{}a(n);~indx~b();~{[}o,nc{]}c()}~
\end{lyxcode}
This signature describes an operator with three parameters \texttt{\small{}a},
\texttt{\small{}b}, and \texttt{\small{}c}. The first, \texttt{\small{}a},
can be any type and has one active dim, called \texttt{\small{}n}.
Your code can loop explicitly over \texttt{\small{}n}, or index particular
values of \texttt{\small{}n} directly. The second, \texttt{\small{}b},
is forced to be an index type and has zero active dims - i.e. your
code will treat it as a scalar. The last, \texttt{\small{}c}, is used
for output (that is the purpose of the \texttt{\small{}o} flag), and
has zero active dims. The \texttt{\small{}nc} flag forces the user
to supply a PDL for output - the code will not autocreate one if it
is missing.

If you have two or more arguments with more than zero active dims,
you can choose whether their active dims are ``the same'' or ``different''
by giving them the same or different names. For example, the signature:
\begin{lyxcode}
{\small{}a(n);~b(n);~{[}o,nc{]}c()}{\small \par}
\end{lyxcode}
describes a function with two arguments and one active dim. The two
arguments must agree on the size of the active dim, (although your
code has full access along that dim).

You can set the exact size of a dimension by assigning to the index
name like a C lvalue in the signature. For example, 
\begin{lyxcode}
{\small{}a(n=3);~{[}o{]}out()}{\small \par}
\end{lyxcode}
specifies a single input argument (\texttt{\small{}a}) with one active
dim (called \texttt{\small{}n}) that has size 3. If the actual passed-in
argument has a different size at runtime, the code will throw an error. 

Sometimes, you might want to calculate a dimension on-the-fly based
on context. You can do that either by conditioning your PP call with
a Perl subroutine (see the paragraph on PMCode, in Section \ref{sub:Basic-pp_def-keys}),
or by using a special hook to generate code that calculates the dims
on the fly (see the paragraph on RedoDimsCode, in Section \ref{sub:More-advanced-pp_def}).

\paragraph{Repeated Dims (Square Operators)}

There is a not-too-uncommon special case where you want to force two
active dims to have the same size (e.g. if your operator works on
a square matrix). For that, you have to specify the same dim name
in two locations, for example:
\begin{lyxcode}
{\small{}a(n,n);~b(n,n)}{\small \par}
\end{lyxcode}
That case is autodetected and works as expected -- \emph{except} that
inside the code block, you refer to the dims with a trailing digit,
to distinguish them -- the first copy is called \texttt{\small{}n0}
and the second is called \texttt{\small{}n1}. The order is the same
in all square variables if you have more than one: the first copy
of \texttt{\small{}n}, working left to right, is referred to as \texttt{\small{}n0};
the second is \texttt{\small{}n1}; etc.

\subsubsection{The Importance of Operator Shape: A Cautionary Example}

It's very important to understand the shape of any operator you intend
to use or build, and to think about how that shape interacts with
the threading engine. For many operators shape is trivial, but for
some it can strongly affect threading behavior and performance. For
example, the built-in \texttt{\small{}index} routine has the signature
\begin{lyxcode}
{\small{}src(n);~indx~dex();~{[}o{]}output()}{\small \par}
\end{lyxcode}
and it carries out lookup by a single number (in \texttt{\small{}dex})
into \texttt{\small{}src}. Provided that \texttt{\small{}src} is 1-D,
or that the user only grabs one value from \texttt{\small{}src}, that
works fine. But a common use case is to thread over additional dims
in \texttt{\small{}src} -- feeding in, say, a 2-D array and selecting
column vectors with \texttt{\small{}dex}. Since both \texttt{\small{}dex}
and \texttt{\small{}src} have thread dimensions, those dimensions
must match. Suppose you want to extract 5 column vectors from a $10\times20$
\texttt{\small{}src} variable. The active dimension \texttt{\small{}n}
gets size 10 and the first thread dim. gets size 20. Since \texttt{\small{}dex}
is declared with no active dims, it must match the first thread dim
- i.e. its dim 0 must have size 20 to match the first thread dim.
Getting your 5 column vectors requires using a $20\times5$ \texttt{\small{}dex}.
(You could supply a $1\times5$ dex, of course, and the thread engine
would automatically extend it to $20\times5$.) That indexing operation
requires 20 times more offset lookups than are strictly necessary,
hurting speed. If you know you are planning to thread both over the
index and over the source file, you can instead use \texttt{\small{}index1d},
which behaves similarly but has the signature:
\begin{lyxcode}
{\small{}src(n);~indx~dex(m);~{[}o{]}output(m)}{\small \par}
\end{lyxcode}
This signature declares the \texttt{\small{}dex} as a 1-D object (to
be treated as a list of indices into the 0 dim of \texttt{\small{}src},
called \texttt{\small{}n}), and declares that the output is also a
1-D object. If \texttt{\small{}src} is a $10\times20$-piddle, and
\texttt{\small{}dex} is a scalar, then the return value will be a
$1\times20$-piddle (\texttt{\small{}where index} would return a 20-piddle),
since the first thread dimension gets size 10. Similarly, if \texttt{\small{}src}
is a $10\times20$-piddle and \texttt{\small{}dex} is a 2-piddle,
then the return value will be a $2\times20$-piddle (where \texttt{\small{}index}
would crash on a thread dim mismatch). In other words, \texttt{\small{}index1d}
threads more appropriately than \texttt{\small{}index,} to pull threaded
vectors out of \texttt{\small{}src}.

That's not to say that \texttt{\small{}index} is inferior to \texttt{\small{}index1d}:
if you plan to thread over only one of the two inputs (either to extract
a single source value for each index, or extract a single column from
source with exactly one index location), \texttt{\small{}index} handles
the output dimensions more conveniently.

\subsubsection{Signature option flags}

You can modify any of your parameters in the Pars field, with one
or more of these options in square brackets:
\begin{description}
\item [{io}] Marks an argument as both input and output. 
\item [{nc}] Marks an argument as necessary (``never create''). This
is the default for normal input parameters; the use case is not obvious,
but I include it here for completeness.
\item [{o}] marks the argument as an output argument; if missing, it will
be created. (If there is only one output argument, it will be returned
as the operator's return value). 
\item [{oca}] marks the argument as an output, and also forces it to be
created (i.e. it should \emph{not} be passed in when the operator
is called). Normal output arguments can be passed in to reuse an existing
variable.
\item [{t}] marks the argument as a temporary variable (not passed in or
out; just created on-the-fly as a scratch space, and erased on return.
The argument is only large enough to hold the active dims; it is overwritten
on each iteration of the surrounding thread loops)
\item [{phys}] Forces the argument to be physicalized -- i.e. calculated
and placed in memory if any calculations are pending, or if the argument
is a complex slice of another PDL. This is necessary if you plan to
access the argument's memory directly, rather than only through the
macros supplied by PP.
\end{description}

\subsubsection{\label{sub:Signature-data-types}Signature data types}

You can specify the data type of a particular argument in the signature.
Normally, a separate version of the code is compiled for each possible
numeric data type, and all arguments are promoted to the ``weakest''
data type that can contain all of them, before your code gets run.
If you specify a type for a particular argument, then it is hammered
into that type regardless of the type of the rest of the arguments. 

PP can accept these type names (in promotion order): \texttt{\small{}byte},
\texttt{\small{}short}, \texttt{\small{}ushort}, \texttt{\small{}int},
\texttt{\small{}indx}, \texttt{\small{}longlong}, \texttt{\small{}float},
and \texttt{\small{}double}. All of these are the same as their basic
C type, except \texttt{\small{}indx} -- which is the type used on
the current system for indexing pointers. (32 or 64 bit signed integer).
You can append a ``+'' to the type to force that argument to take
\emph{at least} the named type or a stronger one. 

If you name the data type of an argument, then that argument is excluded
from the overall data type calculation. So if you specify a signature
like
\begin{lyxcode}
{\small{}float+~a();~b(m,n);~{[}o{]}c}{\small \par}
\end{lyxcode}
then its first argument \texttt{\small{}a} will always be treated
as a \texttt{\small{}float} or \texttt{\small{}double}, but the second
argument \texttt{\small{}b} won't be autopromoted to match the first,
even if you pass in, say, a \texttt{\small{}float} and a \texttt{\small{}byte}
type for \texttt{\footnotesize{}a} and \texttt{\footnotesize{}b} respectively.
If you use the ``+'' format, then the generated code for weaker
argument types will declare your argument as the specified type, but
for stronger types it will be declared as the generic type for the
operation. For example: ``\texttt{\small{}short+ a()}'' will cause
\texttt{\small{}a} to be a \texttt{\small{}short} in the C function
that handles \texttt{\small{}byte} as the generic type, and it will
cause \texttt{\small{}a} to be a \texttt{\small{}float} in the C function
that handles \texttt{\small{}float} as the generic type.

Using data type specification can help cut down on memory overhead
-- for example, specifying a bulky-type temporary variable and manually
promoting data into it inside your function avoids storing the whole
promoted data array.

\subsection{PP definition functions}

Running a PP definition script applies the various definitions in
the file, in the order they appear. The definitions take the form
of calls to functions exported by the PDL::PP module. Each call either
changes the behavior of PP, builds a data structure containing your
code, or exports it as a .pm and .xs file.

The most basic PP file is a Perl script that imports the PDL::PP module
and makes calls to \texttt{\small{}pp\_def()} and \texttt{\small{}pp\_done()}.

\subsubsection{Generating C and XS code}

Here are PP definitions you're likely to find useful, even for basic
stuff.

\paragraph{\texttt{pp\_def} - the core of PP}

\texttt{\small{}pp\_def} accepts an operator name and a hash ref that
describes the operator with several different keywords. It is so important
that it gets its own section below (§\ref{sub:The-pp_def-definition}).

\paragraph{\texttt{pp\_done} - finish a PP definition}

The very last PP call at the bottom of a definition script should
be \texttt{\small{}pp\_done()}. It takes no arguments, and causes
the definitions to be written out to the .xs and .pm files. You can
leave it out of \texttt{\footnotesize{}Inline Pdlpp} scripts, since
\texttt{\footnotesize{}Inline::Pdlpp }puts it in for you.

\paragraph{\texttt{pp\_addpm} - add Perl or POD material to the .pm file (Perl
language)}

PP produces a .xs and a .pm (Perl module) file. \texttt{\small{}pp\_addpm()}
accepts a single string containing Perl language material that gets
inserted into the generated .pm file, in order of pp\_addpm calls.
The added code starts just after the introductory material at the
top of the module, and subsequent calls to \texttt{\small{}pp\_addpm()}
append more lines. It's useful for adding pure-Perl functions or POD.
There are a couple of minor restrictions on the Perl code you can
insert (in particular, don't mess with \texttt{\small{}@EXPORT} or
\texttt{\small{}@ISA} directly; use the definition functions below).

PP keeps track of three different locations in your .pm file -- the
``Top'', ``Middle'', and ``Bottom''. These are handled by three
separate queues of material. By default, \texttt{\small{}pp\_addpm}
adds to the ``Middle'' section, but you can use a two-argument form
to specify which queue you want to add stuff to. Just pass in an initial
argument that is a hash ref -- one of \texttt{\small{}``\{At=>'Top'\}}'',
``\texttt{\small{}\{At=>'Middle'\}}'', or ``\texttt{\small{}\{At=>Bottom'\}}''.
That way you can organize the .pm file more or less independently
of the structure of your .pd file.

\paragraph{\texttt{pp\_addbegin} - add Perl or POD material to a BEGIN block
in the .pm file (Perl language)}

You can add BEGIN block material to your .pm file this way. (You can
also insert an explicit BEGIN block with \texttt{\small{}pp\_addpm}).

\paragraph{\texttt{pp\_addxs} - add extra XS code of your own}

This is handy for adding an extra XS/C function to the generated .xs
file, just like pp\_addpm is useful for adding perl functions to the
generated .pm file. This could be a utility or interface function
that doesn't use threading. You should pass in a string that is a
complete perlXS declaration; it will go in the function-declaration
part of the XS file.

\paragraph{\texttt{pp\_addhdr} - add PerlXS header material (C language)}

This lets you add stuff up in the PerlXS header segment. It comes
{*}after{*} the basic includes and XS setup, so you can define whole
C functions and such in here, as well as just invoking header material.
\texttt{\small{}pp\_addhdr} takes a single argument, which is a string
containing the code you want to add. If you wrap your string in a
\texttt{\small{}pp\_line\_numbers()} call, then any compiler errors
will refer to line numbers in the .pd file (and be much more understandable).

\paragraph{\texttt{pp\_add\_boot} - add XS BOOT code}

This lets you add lines to the declaration/initialization section
of the perlXS file, to set up libraries or initialize globals.

\paragraph{\texttt{pp\_boundsCheck} - turn off/on PDL bounds checking in macros}

By default, the PDL access macros inside your PP code are wrapped
in bounds checks. This normally doesn't cost much time, because the
index variables are usually present in the CPU cache and many processes
are memory-bound anyway -- but you can turn off bounds checking for
a little extra speed (at the risk of dumping core, instead of just
throwing a Perl exception, in case of error). You feed in a single
argument that is evaluated in boolean context. You can check the status
by feeding in no arguments, in which case the current boundsCheck
flag is returned.

\paragraph{\texttt{pp\_line\_numbers}\texttt{\small{} }- insert line number
pragmas into some code}

This is mainly helpful for debugging -- anywhere you hand in code
as a string (mostly in \texttt{\small{}pp\_def}), you can wrap the
string in a \texttt{\small{}pp\_line\_numbers} call to insert line
number pragmas into the code. So instead of saying ``\texttt{\small{}Code=><\textcompwordmark{}<'EOCODE',}''
in your \texttt{\small{}pp\_def} call, you can say \texttt{\small{}``Code
=> pp\_line\_numbers(\_\_LINE\_\_, <\textcompwordmark{}<'EOCODE'),}'',
and your generated code will report the actual line number in the
.pd file instead of a meaningless line number in an intermediate .xs
or .pm file.

\subsubsection{\label{sub:Tweaking-PP-Import/Export}Controlling import/export and
the symbol table}

There are a few PP functions that are specifically for controlling
how modules import and export elements. You're not likely to encounter
or need them for simple tasks but will find them handy if you declare
large modules.

\paragraph{\texttt{pp\_bless}\texttt{\small{} }- set the module into which operators
will be added (default ``PDL'')}

Pass in the name of the module you want to add your operators to.
It defaults to PDL. If you say ``\texttt{\small{}pp\_bless(\_\_PACKAGE\_\_)}''
they will go into the current package from which you are running PP
(probably ``\texttt{\small{}main}'' unless you've declared one).

\paragraph{\texttt{pp\_add\_isa}\texttt{\small{} }- add elements to the module's
\texttt{\small{}@ISA} list}

Just pass in a string containing a module name to be added to \texttt{\small{}@ISA}.
Use this instead of messing with \texttt{\small{}@ISA }directly. If
you call \texttt{\small{}pp\_add\_isa}, you should call \texttt{\small{}pp\_bless}
first, to set the module that you are modifying! This is mainly useful
if/when you're designing a new type of object using PP constructs.

\paragraph{\texttt{pp\_core\_importList} - control imports from PDL::Core}

By default, PP issues a 'use PDL::Core;' line into the .pm file. This
imports the exported-by-default names from PDL::Core into the current
namespace. If you hand in the string ``\texttt{\small{}()}'' you
get none of that, which is handy if you are over-riding one of Core's
methods. Alternatively, you can explicitly list the names you want,
as in ``\texttt{\small{} qw/ barf /} ''.

\paragraph{\texttt{pp\_add\_exported} - export particular Perl functions from
the module}

You supply a single string with a whitespace delimited collection
of Perl function names to export from the module. You should use \texttt{\small{}pp\_add\_exported()}
instead of messing directly with \texttt{\small{}@EXPORT }in a \texttt{\small{}pp\_addpm()}
block. If you don't want to export any function names to the using
package, then don't bother. As with a regular PerlXS module, you can
export both Perl function found in the .pm file and also hybrid compiled
functions named in the .xs file (i.e. things you created with \texttt{\small{}pp\_def}
or \texttt{\small{}pp\_addxs}).

\paragraph{\texttt{pp\_export\_nothing} - don't export defined routines}

By default, PP adds all subs defined using \texttt{\small{}pp\_def}
into the output .pm file's \texttt{\small{}@EXPORT} list. This is
awkward if you don't want them exported into any module that uses
them -- for example if you are creating a subclassed object, or otherwise
have a name conflict. Call \texttt{\small{}pp\_export\_nothing()}
just before \texttt{\small{}pp\_done()}, and nothing will be exported,
keeping the namespace clean

\subsection{\label{sub:The-pp_def-definition}The pp\_def definition function}

Calling \texttt{\small{}pp\_def} is the main way to declare a new
PDL operator. The basic call is simple. You feed in the name of a
function and a collection of hash keys, the values of which are metadata
and/or code for the operator. Inside snippets of code, you can use
special macros to access the data and metadata of the PDL arguments
to your operator. In this section we describe the different keys you
can hand to \texttt{\small{}pp\_def, and }the macros you can use inside
your code sections.

There are two basic forms of operator: calculations (data operations)
and connections (dataflow operations). Calculation operators are straightforward:
they work just like function calls in any other language, returning
zero or more PDLs that are the result of the main calculation but
that do not remain connected to their source. Connections connect
two or more PDLs so that they maintain an ongoing relationship that
is updated automagically -- assigning to or modifying one causes a
recalculation of the other one. In principle connections can perform
calculations (e.g. you can define a ``reciprocal'' operator that
flows data between two PDLs), but in practice only the selection and
slicing operators are connections.

Both types of call use a single ``generic'' data type that is calculated
at call time, from the arguments. Multiple copies of your operator
are compiled, one for each main generic data type, and given slightly
different function names. At call time, all arguments are regularized
to an appropriate generic type and the appropriate copy of your code
(compiled for that generic type) gets called.

\subsubsection{\label{sub:A-simple-PP}A simple pp\_def example: cartND}

Here is a simple example of a PP calculation function, \texttt{\small{}PDL::cartND},
that collapses the 0 dim of its input to find the Cartesian length
of each row vector in it.
\begin{lyxcode}
{\small{}pp\_def('cartND',~~~~~~~~~~~~~~~~~~\#~name~(goes~in~PDL::~package~by~default)}{\small \par}

{\small{}~~~~~~~Pars~=>~'vec(n);~{[}o{]}len;',~\#~Signature~of~the~function}{\small \par}

{\small{}~~~~~~~GenericTypes~=>~{[}'F','D'{]},~\#~Data~types~to~use~-{}-~or~omit~to~use~all}{\small \par}

{\small{}~~~~~~~Code~=>~q\{~~~~~~~~~~~~~~~~~\#~Single~quotes~around~code~block}{\small \par}

{\small{}~~~~~~~~~~\$GENERIC()~acc~=~0;~~~~~\#~Declare~'acc'~to~be~the~generic~type}{\small \par}

{\small{}~~~~~~~~~~loop(n)~\%\{~~~~~~~~~~~~~~\#~Loop~over~the~active~dim~'n'~in~the~signature}{\small \par}

{\small{}~~~~~~~~~~~~acc~+=~\$vec(){*}\$vec();~\#~Parameter~PDLs~are~automagically~indexed~for~you}{\small \par}

{\small{}~~~~~~~~~~\%\}~~~~~~~~~~~~~~~~~~~~~~\#~~{[}outside~loop~you~could~use,~e.g.,~'\$vec(n=>2)'{]}}{\small \par}

{\small{}~~~~~~~~~~\$len~=~sqrt(acc);~~~~~~~\#~Assign~to~the~output~PDL.}{\small \par}

{\small{}~~~~~~~~~\}}{\small \par}

{\small{}~~~~~~);}{\small \par}
\end{lyxcode}
The operator has a name (\texttt{\small{}cartND}), and a collection
of hash values that define the rest of the function to the PP compiler.
The first value has the name \texttt{\small{}Pars} and is the \emph{signature}
of the operator. The operator accepts a single parameter (\texttt{\small{}vec})
with a single active dim, and returns a single parameter (\texttt{\small{}len})
with no active dims. It operates on the data types float and double
only -- its argument will be promoted to float if it begins life as
an integer type. The \texttt{\small{}Code} section is where the basic
operation takes place -- it is written in C with some extra macros,
and gets wrapped automatically with a loop that handles all thread
dims if \texttt{\small{}vec} turns out to have more than one dimension. 

The \texttt{\small{}\$GENERIC()} at the top of the \texttt{\small{}Code}
section is a PP macro -- it expands into a type definition for each
of the data types you list by abbreviation in the \texttt{\small{}GenericTypes}
value. So that line declares \texttt{\small{}acc} to be a holding
variable of whatever type is appropriate based on the arguments' types
at call time.

The \texttt{\small{}loop(n)} and matching \texttt{\small{}\%\{} and
\texttt{\small{}\%\}} form another macro that expands to a C style
\texttt{\small{}for} loop running over all possible values of the
index \texttt{\small{}n}, so the loop accumulates the sum-of-squares
of all the components of the vector. The parameters are accessed with
a leading \texttt{\small{}\$} and a trailing \texttt{\small{}()} pair.
You must specify the value of any active dim indices that aren't explicitly
set or looped over (e.g. with ``\texttt{\small{}\$vec(n=>ii)}''
if you have an index variable \texttt{\small{}ii}) -- but in this
example there's no need. The \texttt{\small{}loop(n)} provides implicit
dereferencing for the \texttt{\small{}n} index.

Note that you don't have to include \emph{any} loops in your PP code
if your signature is scalar -- in that case, each \texttt{\$arg()}
macro stands on its own. (Note that \textbf{you have to include the
parentheses} even if you're not setting the value of any index!)

At the end of the accumulation loop, the \texttt{\small{}Code} takes
the square root of the accumulated value and assigns it to \texttt{\small{}\$len()},
which goes into the output PDL.

That's all!

You can insert the above snippet into a .pd file and compile it, or
into an Inline block in your script and have it compiled on-the-fly.

\subsubsection{\label{sub:Basic-pp_def-keys}Basic pp\_def keys}

Here are a list of common pp\_def keys you need for basic calculations.

\paragraph{\texttt{\small{}Pars} - the operator signature and argument names}

This is the operator signature discussed above (Section \ref{sub:The-signature-of}).
It lists all the PDL arguments accepted by your operator, and their
active dims (if any), access modes, and allowable type(s) for that
argument (if different from the generic type). An example for a scalar
binomial operator is ``\texttt{\small{}Pars => 'a(); b(); {[}o{]}out();'}''.
Another example, for outer product of two vectors, is ``\texttt{\small{}Pars
=> 'a(n); b(m); {[}o{]}out(m,n)'}''.

\paragraph{\texttt{\small{}OtherPars} - non-PDL argument names}

After all the PDL arguments to a PP function you can also pass in
other parameters that are not PDLs. These are handled with the OtherPars
section. These can be declared as C types or Perl C-side types (including
\texttt{\small{}SV {*}}) and will be converted on-the-fly from their
scalar values on the Perl side of the operator call.\emph{ }A caveat
is that you can't easily provide a custom typemap for these arguments.
If you have, for example, a structure pointer to pass in you can pass
it in as a Perl \texttt{\small{}IV} and cast it to a pointer when
you use it (PDL's interface to the Gnu Scientific Library does that).
Alternatively, you can pass in pretty much anything as an \texttt{\small{}SV
{*}} and allow your \texttt{\small{}Code} section to handle it directly.
An example OtherPars is ``\texttt{\small{}OtherPars=>'double foocoeff;
IV foocount; SV {*} perlsv;'}''. 

Note that, even if you pass in lots of OtherPars, you must always
pass in at least one PDL argument in the Pars section, or PP will
choke.

The OtherPars are not available as directly named variables or direct
macros. You get them via a macro called \texttt{\small{}\$COMP()}
that is accessible in the Code segment. For example, with the above
declaration, you could access foocoeff by saying ``\texttt{\small{}\$COMP(foocoeff)}''.
COMP stands for ``Compiled'', and it is a data structure compiled
by PP to hold ancillary data during setup for an operation.

\paragraph{\texttt{\small{}PMCode} - a Perl-side handler}

PMCode lets you define some Perl code that gets executed instead of
the \texttt{\small{}Code} block when your operator gets called. There
could be several reasons. For example, you might want to precondition
the arguments to your PP function, or want to be sure you don't break
dataflow connections by autoconverting the arguments to your function;
you might want to prepare default values for optional arguments to
the main Code block; or you may even want to mock up the Code block
in Perl before implementing it. \texttt{\small{}PMCode} lets you handle
those cases. 

If you want to use your Perl code to condition the arguments for the
compiled code, you can include a \texttt{\small{}Code} block \textbf{and}
a \texttt{\small{}PMCode} block. That will generate both types of
code, but only the \texttt{\small{}PMCode} gets executed directly.
Your PMCode can call the special function ``\texttt{\small{}\&}\texttt{\emph{\small{}<package>}}\texttt{\small{}::\_}\emph{<name>}\texttt{\small{}\_int()}''
with the arguments given in Pars and OtherPars to run the \texttt{\small{}Code}
block. The \emph{<package> }is either ``\texttt{\small{}PDL}'' by
default, or the value you specified with a call to pp\_bless(). That
mechanism lets you do some initial parsing or argument currying with
Perl, then jump into your compiled code for faster processing of the
algorithmic ``hot spot''. \emph{Note: This functionality does not
work for Inline PP modules, for all versions of PDL through 2.011.
That is a ``known problem'' that may be fixed in a future release
of PDL. To curry arguments with inline code, you must (for all versions
at least through 2.011) declare a ``front-end'' perl function and
a separate PP operator with no }\texttt{\emph{\small{}PMCode}}\emph{
block.}

\paragraph{\texttt{\small{}Code} - the main code to execute}

The compiled code you want to implement. See Section \ref{sub:pp_def-Code-macros}
for details on what goes in here. This is the main Code block, but
there can be several code blocks even in a simple calculation function
(see \texttt{\small{}BadCode} and \texttt{\small{}PMCode}, below).
You don't actually need to declare a Code block at all if your operator
is written entirely in Perl (you can use the \texttt{\small{}PMCode}
key for that).

\paragraph{\texttt{\small{}BadCode} - A variant of \texttt{\small{}Code} }

PDL can handle bad values in data. A particular in-band value (e.g.
-32768 for shorts) is chosen to mark data as bad or missing. Bad values
are intended to be contagious. Handling them requires additional logic.
For example, \texttt{\small{}``\$c() = \$a() + \$b()}'' becomes
something like:
\begin{lyxcode}
{\small{}if(~!\$ISBAD(\$a())~\&\&~!\$ISBAD(\$b())~)}{\small \par}

{\small{}~~~\$c()~=~\$a()~+~\$b();}{\small \par}

{\small{}else~}{\small \par}
\begin{lyxcode}
{\small{}\$c()~=~BADVAL;}{\small \par}
\end{lyxcode}
\end{lyxcode}
which requires a few more CPU cycles than the direct addition. Specifying
a separate BadCode segment lets PP switch between direct operations
and their bad-value-handling variants based on whether the arguments'
badflags are set (i.e. whether bad values may be present or cannot
possibly be present). See Section \ref{sub:BAD-values} for details.

\paragraph{\texttt{\small{}GenericTypes} - set specific types to be compiled}

PP normally creates a separate version of your operator for each data
type recognized by PDL. When the routine is called by name, the PDL
threading engine reconciles the variable types the same way they would
be reconciled in an arithmetic expression -- and the call is dispatched
to the appropriate version of the operator (in which the argument
types are declared appropriately, and the \texttt{\small{}\$GENERIC()}
macro expands to that type declaration in C). But sometimes you want
to support only a subset of the available types. You pass in a Perl
array ref containing the single-letter specifiers for the types you
want to support: 'B' for byte, 'S' for short, 'L' for long, 'N' for
index (32 or 64 bits as appropriate to your system), 'F' for float,
and 'D' for double. Unrecognized types will yield a compiler error.

\paragraph{\texttt{\small{}Inplace} - Set inplace handling}

If this is set to be true, then your routine can handle in-place processing
of its arguments (to save memory, or for convenience). When an argument
is processed in-place, its result is placed back into the original
piddle instead of copied to an output variable. The in-place operation
is handled fully by the PP engine, including clearing the inplace
flag after the operation completes. If you want to handle in-place
operation manually, you can check the flag for yourself by masking
the parameter's state field with the constant \texttt{\small{}PDL\_INPLACE}
(declared in \texttt{\small{}pdl.h}) -- but remember to clear that
flag before returning!

If your operator has only one input and only one output, you can set
``\texttt{\small{}Inplace=>1}'', and the input and output variable
will be linked if the input has its \texttt{\small{}inplace} flag
set. If you have more inputs and/or more outputs, you must specify
a two-element array ref with the names of the input and output parameter
to link. For example: ``\texttt{\small{}Pars=>'a(); z(); {[}o{]}b();',
Inplace=>{[}'a','b'{]}}''. Of course, the two parameters must have
the same signature in the \texttt{\small{}Pars} declaration, except
that the \texttt{\small{}{[}o{]}} modifier must be set for the second. 

You can link at most one input to one output parameter using the built-in
inplace handling, though of course you can roll your own by accessing
the \texttt{\small{}inplace} flag of each argument.

\paragraph{\texttt{\small{}HandleBad} - set bad value handling}

If you set \texttt{\small{}HandleBad=>0}, then any input argument
with its BadFlag set will cause an exception. If you set it to \texttt{\small{}undef},
or don't set it at all, then bad values are ignored and treated as
any other value.

If you set \texttt{\small{}HandleBad=>1}, and declare \texttt{\small{}BadCode}
(below), then the BadCode will get run whenever it's appropriate (i.e.
one or more input PDLs has its BadFlag set and therefore might contain
specially coded missing values). Also, PP will define all the relevant
bad-value handling macros \texttt{\small{}\$ISBAD()}, \texttt{\small{}\$SETBAD()},
etc. See Section \ref{sub:BAD-values}.

\paragraph{\texttt{\small{}Doc} - POD documentation}

You can document your project using POD, Perl's Plain Old Documentation
format. You can in principle add POD using a PMCode block (since POD
is a valid Perl no-op), but using the Doc field will cause your project
to be indexed into the system PDL documentation database if/when you
``make install'' your module.

\paragraph{\texttt{\small{}BadDoc} - Special POD for bad value handling}

If the installed version of PDL was compiled to handle bad values,
this documentation gets appended to the documentation in the \texttt{\small{}Doc}
field.

\paragraph{\texttt{\small{}NoPThread} - Mark Function as not threadsafe}

By default, PP reserves the right to run multiple instances of your
code in multiple CPU threads, for extra speed on multiprocessor systems.
(If it does this, temporary variables in the signature are properly
duplicated across threads). If your code is not threadsafe (for example,
if you write to local variables or the \texttt{\small{}\$COMP} structure
without your own MUTEX), you can set this flag to avoid multithreading.
(Note that multithreading is currently a \emph{language-compile-time
experimental option} in PDL 2.009, but it may be enabled by default
in a future version.)

Because the nomenclature is confusing, here's a reminder. A PThread
is a Perl thread, which is generally implemented as a CPU thread.
A CPU thread is an execution environment with a unique program counter
-- unlike a CPU process, it shares memory with other threads in the
same execution context. A PThread is a CPU-like thread that exists
in some Perl environments, depending on how \texttt{\footnotesize{}perl}
was compiled. These types of thread are distinct from PDL's ``threading'',
which is vectorization over thread dims of a PDL. A MUTEX is a mutual-exclusion
gate that prevents multiple threads from writing to the same memory
simultaneously, or prevents one from reading from a program variable
while a different one is writing to it.

\paragraph{\texttt{\small{}PMFunc} - Control Where the Function gets Linked}

Normally your PP code gets compiled and linked into the PDL module
(or whichever module you specified with \texttt{\small{}pp\_bless},
if you used that - see Section \ref{sub:More-advanced-pp_def}). PMFunc
adds a hard link to the function name of your choice. It's functionally
equivalent to adding a line ``\texttt{\small{}{*}}\emph{name}\texttt{\small{}=\textbackslash{}\&PDL::}\emph{name}\texttt{\small{};}''
to the Perl part of your definition. It appears, as of PDL 2.009,
to not work with inline declarations, though it works with module
declarations.

\paragraph{\texttt{\small{}RedoDimsCode} - Calculate the Dimensionality of the
Output at Runtime}

This field isn't normally necessary, but you can use it to calculate
the dimensionality of an output variable at runtime. That's useful
for some applications where the output variable's shape is related
in a non-simple way to the input variables'. See Section \ref{sub:Adjusting-output-dimensions}
for details.

\subsubsection{\label{sub:More-advanced-pp_def}More advanced pp\_def keys}

You can pass in many more keys to pp\_def than the basics described
above. The most important of these are associated with dataflow. They
are covered in Chapter \ref{sec:Using-PP-With}.

\subsubsection{\label{sub:pp_def-Code-macros}pp\_def Code macros}

In the C code blocks like \texttt{\small{}Code} or \texttt{\small{}BadCode},
you write in C -- with preprocessing. The PDL ``hooks'' are implemented
with text replacement macros that give you the functionality you need.
The basic ones are below. You can find other macros for handling bad
values, in Section \ref{sub:BAD-values}.

\paragraph{\texttt{\small{}\$COMP(}\emph{name}) - access the compiled data structure
and OtherPars parameters}

\texttt{\small{}\$COMP} accesses a compiled data structure associated
with the operator. In computation operators, \texttt{\small{}\$COMP}
is by default pre-loaded with all (and only) the non-threading parameters
you declare with an \texttt{\small{}OtherPars} field in \texttt{\small{}pp\_def}.
You can declare additional \texttt{\small{}\$COMP} fields with the
\texttt{\small{}Comp} argument to \texttt{\small{}pp\_def}, and/or
perform additional computations up-front to generate \texttt{\small{}\$COMP}
values with a \texttt{\small{}MakeComp} block (See Section \ref{sub:More-advanced-pp_def}).
That's primarily useful in dataflow operators. The \texttt{\small{}\$COMP(}\emph{name}\texttt{\small{})}
C fields are lvalues.

\paragraph{\texttt{\small{}\$}\emph{name}\texttt{\small{}()} - access elements
in threaded Pars parameters}

To access the value in one of the parameters named in the \texttt{\small{}Pars}
signature of your operator, you just name it, with a prepended \texttt{\small{}\$}
and a postpended pair of parentheses. (Note that this means you should
not name a parameter, e.g., ``COMP''!) If there is a named dimension
in the parameter, you must index that dimension with a double-arrow,
as in \texttt{\small{}\$a(n=>2)} for a parameter \texttt{\small{}a}
with an active dimension called \texttt{\small{}n}. The \texttt{\small{}loop}
construct, below, automagically selects the index in the dimension
being looped over, so you don't have to. If you have to specify multiple
indices, you can separate them with commas, as in \texttt{\small{}\$a(n=>2,m=>3)}
to specify location in a variable with two active dims. 

If a dim appears twice in the signature, as in ``\texttt{\small{}a(n,n);}''
in the Pars section to specify a 2-D square variable, then you access
the two dimensions by appending a numeral, starting at 0, as in ``\texttt{\small{}\$a(n0=>2,n1=>3)}''.
The numerals start fresh at 0 for each Par with a square (or greater)
aspect. There's currently no way to transpose two square parameters
relative to one another: the first copy of the repeated name in each
parameter maps to \emph{name}\texttt{\small{}0} for all square parameters,
and the second maps to \emph{name}1.

\paragraph{\texttt{\small{}\$SIZE(}\emph{iname}\texttt{\small{})} - return the
size of a named dimension}

You often need to know the size of a named dimension. This macro expands
to it.

\paragraph{\texttt{\small{}\$GENERIC()} - the generic C type}

\texttt{\small{}\$GENERIC()} expands to a C type declaration for the
generic type of the operation. All incoming PDL parameters' types
will be resolved to the smallest type that can represent all of them,
and converted to that type. So you can declare an internal working
variable \texttt{\small{}x}, for example, as ``\texttt{\small{}\$GENERIC()
x=0;}'' in your \texttt{\small{}Code} block. PP will compile a separate
copy of your code for each possible value of \texttt{\small{}\$GENERIC()},
and dispatch execution accordingly at runtime.

\paragraph{\texttt{\small{}\$T}\emph{foo}\texttt{\small{}()} - type-dependent
syntax}

The \$T macro is used to switch syntax according to the generic type.
Each generic type has a one-letter abbreviation: ``B'' for byte,
``U'' for unsigned short, ``S'' for short, ``L'' for long, ``I''
for Indx, ``F'' for float, and ``D'' for double. You string the
letters together after the \$T, and give the macro a comma-separated
list of words to insert in the code depending on the value of the
generic type. For example, ``\texttt{\small{}a = ( \$TSLFD( short,
long, float, double ) ) b;}'' is a limited equivalent to ``\texttt{\small{}a
= (\$GENERIC()) b;}''. \$T is useful for dispatching calls to a type-specific
library function, or any other situation where you need to make a
small difference between each of the generic-type copies of the compiled
code. \emph{Note: replacement strings containing commas are not supported!}

\paragraph{\texttt{\small{}\$PDL(}\emph{\small{}var}\texttt{\small{})} - return
a pointer to the PDL struct}

On occasion you need to dive into the guts of the C struct associated
with a particular PDL (see §\ref{sub:The-struct-pdl}). This macro
returns a C pointer of type ``\texttt{\small{}struct PDL {*}}'',
so you can access the internals directly.

\paragraph{\texttt{\small{}threadloop \%\{~}{\small{}...}\texttt{\small{}~\%\}}
- explicitly locate the thread loop}

PP autogenerates looping code to repeat your operation for each element
along thread (extra) dimensions in the input parameters. By default
your entire \texttt{\small{}Code} block is run separately for each
threaded-over element. For some types of operation this is inefficient
-- for example, you might carry out some initialization that applies
to each threaded instance of the operation. If you explicitly place
your code between the \texttt{\small{}\%\{} and \texttt{\small{}\%\}}
of a threadloop macro, the thread looping code will be placed there
only.

\paragraph{\texttt{\small{}loop(}\emph{iname}{\small{})~}\texttt{\small{}\%\{~}...{\small{}~}\texttt{\small{}\%\}}
- explicitly loop over a named dimension}

This operates like threadloop, but generates an explicit loop over
just the named dimension. If you use a parameter macro inside the
loop, you do not need to specify the value of the named dimension
index. You can override the loop by specifying it anyway. Of course,
you are also free to generate loop constructs explictly with temporary
index variables and normal C looping structures, but loop can be more
convenient and readable.

\section{\label{sec:Putting-it-together:}Putting it together: Example PP
functions}

Declaring a PP function is only a part of the process of compilation.
Here are examples of how to create a standalone PP module and how
to include PP code inline in your Perl code. You can use these examples
as templates for your own PP projects.

\subsection{\label{sub:An-Inline::Pdlpp-Call}Using PP inline with Perl, via
Inline::Pdlpp}

\texttt{\small{}Inline::Pdlpp} is terrific for placing a compiled
``hot spot'' function right in the middle of your Perl script. We
already saw a simple example in §\ref{sub:An-early-example:}. Here
are some more examples: a simple one that generates Spirograph$^{\textrm{TM}}$-like
parameterized hypocycloid curves, and a more complex one that enumerates
a fractal path through a 2-D grid.

There is an important wart with Inline::Pdlpp compared to a full module
definition: at least through PDL 2.008, the \texttt{\small{}PMCode}
and field for \texttt{\small{}pp\_def} doesn't work correctly. It
is simply ignored by the \texttt{\small{}Inline} compilation process.
If you need a perl helper to precondition inputs, you need to write
it explicitly and call the PP function from there. 

You can use a call to pp\_\texttt{\small{}bless} to place the helper
function in the package you want, but it's also common to just use
a single \texttt{\small{}pp\_def} call and let PP compile the helper
into the default PDL package (as in the examples here).

\subsubsection{\label{sub:A-simple-case:}A simple case: combining compound math
operations }

This example implements \texttt{\small{}spiro}, a compound math function
that generates a parameterized cyclocycloid in 2-D from a single time
parameter and some coefficients. Compound math functions like \texttt{\small{}spiro}
are good candidates for PP, because they require many atomic math
functions to implement -- which breaks cache if you thread over a
large dataset. By merging the operations (in C/PP) \emph{inside} the
thread loop, you can make spiro work up to 10x faster on large data
sets and multiple elementary math operations. An advantage of Inline::Pdlpp
is that you can prototype your code in PDL, then upgrade it in-place
later in exactly the same location, and call it in exactly the same
way. You do not have to set up a module file structure -- Inline handles
all the compilation behind the scenes.

Here are both a PDL and a PP version of \texttt{\small{}spiro}. You
could place them as part of a larger module declaration, or paste
them into a Perl script to be defined in-place, or place them in autoloader
files.

The input \texttt{\small{}\$s} parameterizes length along the spiral.
\texttt{\small{}\$r1} and \texttt{\small{}\$r2} are the sizes of the
outer and inner ``gears'' in the cyclocycloid apparatus, and \texttt{\small{}\$n1
}and \texttt{\small{}\$n2} are the number of teeth. 

The code below implements the same operation in regular PDL idiom,
and also in PP using Inline -- you can place both definitions together
in a single script (or, for that matter, module). Using PP moves memory
accesses inside the loop, preserving cache and speeding up the code
by \textasciitilde{}3$\times$ for moderate-sized curves of up to
a million points. 
\begin{lyxcode}
{\small{}=head2~spiro~-~parameterized~spirograph-like~cyclocycloid}{\small \par}

~

{\small{}=for~usage}{\small \par}

{\small{}~}{\small \par}

{\small{}~\$xy~=~spiro\_pdl(\$s,~\$r1,~\$n1,~\$r2,~\$n2);}{\small \par}

{\small{}~\$xy~=~spiro\_pp(\$s,~\$r1,~\$n1,~\$r2,~\$n2);}{\small \par}

{\small{}~}{\small \par}

{\small{}=cut}{\small \par}

{\small{}sub~spiro\_pdl~\{}{\small \par}
\begin{lyxcode}
{\small{}my(\$s,\$r1,\$t1,\$r2,\$t2)~=~@\_;}{\small \par}

{\small{}my~\$phase1~=~(\$s/\$r2);}{\small \par}

{\small{}my~\$phase2~=~\$phase1~{*}~\$t1/\$t2;}{\small \par}

{\small{}my~\$xy1~=~pdl(cos(\$phase1),~sin(\$phase1))->mv(-1,0)~{*}~\$r1;}{\small \par}

{\small{}my~\$xy2~=~pdl(cos(\$phase2),~sin(\$phase2))->mv(-1,0)~{*}~\$r2;}{\small \par}

{\small{}return~\$xy1~+~\$xy2;}{\small \par}
\end{lyxcode}
{\small{}\}}{\small \par}

{\small{}no~PDL::NiceSlice;}{\small \par}

{\small{}use~Inline~Pdlpp~=>~<\textcompwordmark{}<'EOF';~}{\small \par}

{\small{}pp\_def('spiro\_pp',}{\small \par}
\begin{lyxcode}
{\small{}Pars=>'s();~r1();~t1();~r2();~t2();~{[}o{]}xy(n=2)',}{\small \par}

{\small{}Code=>~q\{}{\small \par}
\begin{lyxcode}
{\small{}double~ph1~=~\$s()~/~\$r2());~}{\small \par}

{\small{}double~ph2~=~ph1~{*}~\$t1()~/~\$t2();}{\small \par}

{\small{}\$xy(n=>0)~=~cos(ph1)~{*}~\$r1()~+~cos(ph2)~{*}~\$r2();}{\small \par}

{\small{}\$xy(n=>1)~=~sin(ph1)~{*}~\$r1()~+~sin(ph2)~{*}~\$r2();}{\small \par}
\end{lyxcode}
{\small{}\}~);}{\small \par}
\end{lyxcode}
{\small{}EOF}{\small \par}

{\small{}{*}spiro\_pp~=~\textbackslash{}\&PDL::spiro\_pp;~\#~copy~to~current~package}{\small \par}

{\small{}print~``Loaded~spiro\_pdl~and~spiro\_pp!\textbackslash{}n'';}{\small \par}
\end{lyxcode}
The two versions of \texttt{\small{}spiro} have very similar structure.
Both versions accept a collection of points to plot (the ``s'' parameter),
two gear tooth counts, and two gear radii. They calculate the corresponding
2-D location of the cyclocycloid at the given ``s''. The Perl version
uses threaded arithmetic operations to assemble the output. The PP
version does the same calculation in C -- but, more importantly, the
C calculation is performed \emph{inside} the threadloop, so that all
of the relevant variables can be loaded into registers and/or CPU
cache.

The signature of \texttt{\small{}spiro\_pp} shows that it is a scalar
operator with 5 inputs. The output has 1 active dim, of size 2 --
because it is a collection of 2-D vectors. The arithmetic operations
are exactly the same as in \texttt{\small{}spiro\_pdl}, translated
to C -- except that the assignments and calculations in \texttt{\small{}spiro\_pp}
are normal C language scalar operations.

\subsubsection{Calling a C function with Inline::Pdlpp}

The\texttt{\small{} hilbert} routine demonstrates jumping out to a
C routine to get useful things done. It calculates an approximation
to the Hilbert curve, which is a linear fractal that fills the plane
(it has Hausdorff dimension 2). The Hilbert curve is useful for dithering
greyscale images to a single bit, or other applications that require
traversing a 2-D grid of locations without rasterizing them. The Hilbert
curve, being fractal, is infinitely complex -- so \texttt{\small{}hilbert}
is recursive, producing an approximation that has been refined through
a specified number of levels of recursion.

The C code here populates a string with glyphs that indicate which
direction the Hilbert curve steps next; it is recursive and descends
\emph{level} steps to refine the curve. The PP code parses the returned
string and uses it to populate a collection of 2-vectors describing
the vertex locations of the curve. For convenience, I've interspersed
explanatory text with code snippets -- but all the code in this section
can be pasted into a single file for the PDL AutoLoader.
\begin{lyxcode}
{\small{}=head2~hilbert~-~generate~a~Hilbert~ordering~of~planar~points}{\small \par}

{\small{}=for~usage~}{\small \par}

{\small{}~~\$line~=~hilbert(\$pdl);}{\small \par}

{\small{}~~\$line~=~hilbert(\$w,\$h);}{\small \par}

{\small{}=for~ref}{\small \par}

{\small{}The~Hilbert~curve~fills~the~unit~square~with~a~linear~area-filling~fractal~curve.~~This~routine~uses~approximations~to~the~Hilbert~curve~to~order~the~integer~2-D~locations~from~0~to~\$w-1~and~0~to~\$h-1.~~The~resulting~ordering~of~the~pixel~plane~is~better~than~rastering~for~some~purposes.~}{\small \par}

{\small{}=cut}{\small \par}
\end{lyxcode}
This top part of the code is just POD that you might put in any .pdl
script.
\begin{lyxcode}

{\small{}\#\#\#\#\#~Here~is~the~main~entry~point~-{}-~a~Perl~subroutine~that~allocates~the}{\small \par}

{\small{}\#\#\#\#\#~output~array~and~jumps~into~the~PP~code.}{\small \par}

{\small{}sub~hilbert~\{~}{\small \par}
\begin{lyxcode}
{\small{}my~\$w,~\$h;~}{\small \par}

{\small{}(\$w,\$h)~=~(ref~\$\_{[}0{]}~eq~'PDL')~?~shift->dims~:~splice(@\_,02);}{\small \par}

{\small{}my~\$dim~=~(~\$w>\$h~?~\$w~:~\$h~);}{\small \par}

{\small{}\$n~=~(log(pdl(\$dim))/log(2))->ceil;~\$siz~=~2~{*}{*}~\$n;}{\small \par}

{\small{}\$coords~=~PDL->new\_from\_specification(long,~2,~\$siz{*}\$siz);}{\small \par}

{\small{}PDL::hpp(\$coords,\$n);~}{\small \par}

{\small{}return~\$coords;~}{\small \par}
\end{lyxcode}
{\small{}\}}{\small \par}
\end{lyxcode}
The actual perl function being defined is \texttt{\small{}hilbert}.
That Perl function calls a PP function declared below -- \texttt{\small{}PDL::hpp}
-- which, in turn, calls a C function. \texttt{\small{}PDL::hpp} gets
called with a single output PDL (\texttt{\small{}\$coords}) that has
the $2\times size^{2}$ dimensions calculated above. Notice that \texttt{\small{}\$coords}
is defined with \texttt{\small{}PDL->new\_from\_specification()},
rather than (say) \texttt{\small{}zeroes()}. That eliminates one pass
through the memory, since the values in \texttt{\small{}\$coords}
are not actually initialized -- they get whatever happened to be sitting
in memory. 

Here is the actual inline code. The ``\texttt{\small{}no PDL::NiceSlice}''
is necessary because NiceSlice interferes with C code parsing. Everything
between the ``\texttt{\small{}use Inline Pdlpp}'' and the ``\texttt{\small{}EOF}''
below is fed to PP to generate a .pm and .xs file that are then linked
in on-the-fly. The code defines a C function and a PP interface function. 
\begin{lyxcode}
{\small{}no~PDL::NiceSlice;~}{\small \par}

{\small{}use~Inline~Pdlpp=><\textcompwordmark{}<'EOF';~\#~exactly~like~this~(no~'::')}{\small \par}
\end{lyxcode}
This is the first PP call (inside the PP block). It is a declaration
of a C utility routine. Here is where you could \texttt{\small{}\#include}
external libraries, for example, if you need them. We use \texttt{\small{}pp\_addhdr}
to put this piece of C code at the top of the generated code.
\begin{lyxcode}
{\small{}pp\_addhdr(~<\textcompwordmark{}<~'EOAHD');~}{\small \par}

{\small{}PDL\_COMMENT(``~This~stuff~goes~below~the~top-of-file~declarations~and~~~'')~}{\small \par}

{\small{}PDL\_COMMENT(``~before~the~XS~stuff,~in~the~compiled~C~code.~It's~a~good~'')}{\small \par}

{\small{}PDL\_COMMENT(``~place~to~declare~a~helper~C~function.~~~~~~~~~~~~~~~~~~~~'')}{\small \par}

{\small{}static~char~{*}hre(char~{*}p,~char~dir,~int~level)~\{~}{\small \par}
\begin{lyxcode}
{\small{}if(level==0)~\{}{\small \par}
\begin{lyxcode}
{\small{}switch(dir)~\{}{\small \par}
\begin{lyxcode}
{\small{}case~'<':~strcpy(p,''>v<'');~break;~~~case~'>':~strcpy(p,''<\textasciicircum{}>'');~break;~}{\small \par}

{\small{}case~'\textasciicircum{}':~strcpy(p,''v>\textasciicircum{}'');~break;~~~case~'v':~strcpy(p,''\textasciicircum{}<v'');~break;~}{\small \par}
\end{lyxcode}
{\small{}\}~~}{\small \par}

{\small{}p~+=~3;}{\small \par}
\end{lyxcode}
{\small{}\}~else~\{~}{\small \par}
\begin{lyxcode}
{\small{}int~l~=~level~-~1;~}{\small \par}

{\small{}switch(dir)~\{~}{\small \par}
\begin{lyxcode}
{\small{}case~'<':~p=hre(p,'\textasciicircum{}',l);~{*}(p++)='>';~p=hre(p,'<',l);~{*}(p++)='v';~}{\small \par}

{\small{}~~~~~~~~~~p=hre(p,'<',l);~{*}(p++)='<';~p=hre(p,'v',l);~break;~}{\small \par}

{\small{}case~'>':~p=hre(p,'v',l);~{*}(p++)='<';~p=hre(p,'>',l);~{*}(p++)='\textasciicircum{}';~}{\small \par}

{\small{}~~~~~~~~~~p=hre(p,'>',l);~{*}(p++)='>';~p=hre(p,'\textasciicircum{}',l);~break;~}{\small \par}

{\small{}case~'\textasciicircum{}':~p=hre(p,'<',l);~{*}(p++)='v';~p=hre(p,'\textasciicircum{}',l);~{*}(p++)='>';~}{\small \par}

{\small{}~~~~~~~~~~p=hre(p,'\textasciicircum{}',l);~{*}(p++)='\textasciicircum{}';~p=hre(p,'>',l);~break;~}{\small \par}

{\small{}case~'v':~p=hre(p,'>',l);~{*}(p++)='\textasciicircum{}';~p=hre(p,'v',l);~{*}(p++)='<';~}{\small \par}

{\small{}~~~~~~~~~~p=hre(p,'v',l);~{*}(p++)='v';~p=hre(p,'<',l);~break;~}{\small \par}
\end{lyxcode}
{\small{}\}~}{\small \par}
\end{lyxcode}
{\small{}\}~}{\small \par}

{\small{}return~p;~}{\small \par}
\end{lyxcode}
{\small{}\}~}{\small \par}

{\small{}EOAHD}{\small \par}
\end{lyxcode}
The EOAHD ended the first PP call (we're still inside the PP definition
block), so we're ready to call \texttt{\small{}pp\_def}. There is
a single PDL parameter in the \texttt{\small{}Pars} signature - a
collection of vectors that we will populate. It's declared ``nc''
for clarity, because PP can't autocreate it (it would have no way
of knowing the size of the named dimension \texttt{\small{}m}) The
named dimension \texttt{\small{}n} is forced to have size 2, since
it indexes an (x,y) pair for each of the \emph{m} points to be reported.
The tag points out that we \emph{only} want a single copy of the code,
suitable for running on the \texttt{\small{}long} type. One non-PDL
parameter is accepted -- the recursion level to which \texttt{\small{}hre}
should descend. It's autoconverted to i\texttt{\small{}nt} by PP and
the PerlXS mechanism. The \texttt{\small{}level} parameter gets accessed
with the \texttt{\small{}\$COMP(level)} macro.

The \texttt{\small{}hre} code needs a character buffer in which to
put the motion directions, so we allocate that. The size of each active
dim is available through the \texttt{\small{}\$SIZE() }macro, and
we use that to make the buffer the correct size (the size was precomputed
by \texttt{\small{}hilbert} before entry).

The meat of the code is simple: we call the C routine that does the
work, then convert its recursively-generated step code into a sequence
of coordinates of the vertices of the curve. Because \texttt{\small{}n}
and \texttt{\small{}m} are declared active dimensions, they are not
automatically looped over -- our code has to do that. But we don't
have to use a \texttt{\small{}for} loop - the \texttt{\small{}loop(m)}
macro takes care of that. Take care to use the macro brackets \texttt{\small{}\%\{}
and \texttt{\small{}\%\}}, rather than simple brackets! Inside the
loop, the variable \texttt{\small{}m}, with no modifiers, is predeclared
and has the correct value. We access \texttt{\small{}out} with the
macro \texttt{\small{}\$out()}. Inside the macro parens, we have to
give values to all the active dimensions that are not already defined.
Since we're in the \texttt{\small{}loop(m)} construct, \texttt{\small{}m}
is defined -- but \texttt{\small{}n} is not. The \texttt{\small{}\$out()}
macro is an lvalue, and we assign the accumulated x and y values to
it. 

The \texttt{\small{}loop(m)} loop is nested inside a construct called
\texttt{\small{}threadloop}. That is important in case of threading
-- by default, PP wraps the threadloop around the whole code -- but
we don't want to call \texttt{\small{}hre()} once for each hyperplane,
if the user decides to thread his call. We only need to call it once.
Placing the threadloop macro around the hotspot lets the code call
\texttt{\small{}hre()} just once (at the start) and then get down
to the business of threading over higher dimensions. The \texttt{\small{}threadloop}
macro can even be placed inside the other loop -- we could in principle
save even a little more time by putting the \texttt{\small{}loop(m)}
macro on the outside and nesting only the final two assignments inside
a tiny \texttt{\small{}threadloop}. 
\begin{lyxcode}
{\small{}pp\_def('hpp',~}{\small \par}

{\small{}~~~~~~~Pars=>'{[}o,nc{]}out(n=2,m);',~\#~Note:~must~be~passed~in~to~define~m~size!}{\small \par}

{\small{}~~~~~~~GenericTypes=>{[}L{]},~}{\small \par}

{\small{}~~~~~~~OtherPars=>'int~level',~}{\small \par}

{\small{}~~~~~~~Code=>~<\textcompwordmark{}<'EOHPP'~}{\small \par}

{\small{}~~~~~long~dex,~side,~x,~y;~}{\small \par}

{\small{}~~~~~char~{*}s,~{*}buf~=~(char~{*})malloc(\$SIZE(m)+2);}{\small \par}

{\small{}\$PDL\_COMMENT(``~hre~is~the~expensive~computation.~Since~it's~exactly~the~same~~~``);}{\small \par}

{\small{}\$PDL\_COMMENT(``~for~all~operations~of~the~same~size,~we~call~it~{*}outside{*}~an~~~~``);}{\small \par}

{\small{}\$PDL\_COMMENT(``~explicit~threadloop.~~That~way~it~only~gets~executed~once,~even~``);}{\small \par}

{\small{}\$PDL\_COMMENT(``~for~a~threaded~operation.~~~~~~~~~~~~~~~~~~~~~~~~~~~~~~~~~~~~~~~``);}{\small \par}

{\small{}~~~~~hre(buf,~'>',~\$COMP(level)-1);~}{\small \par}

{\small{}~~~~~s~=~buf;}{\small \par}
\begin{lyxcode}
{\small{}threadloop~\%\{~~~~~\#~Threadloop~specifies~the~``hot~spot''~for~threaded~operations.~}{\small \par}

{\small{}~~~~~loop(m)~\%\{~}{\small \par}

{\small{}~~~~~~~~if(!m)\{~x=y=0;~\}~else~\{~}{\small \par}

{\small{}~~~~~~~~~~~switch(~{*}(s++)~)\{~}{\small \par}

{\small{}~~~~~~~~~~~~~case~'\textasciicircum{}':~y++;~break;~}{\small \par}

{\small{}~~~~~~~~~~~~~case~'v':~y-{}-;~break;~}{\small \par}

{\small{}~~~~~~~~~~~~~case~'<':~x++;~break;~}{\small \par}

{\small{}~~~~~~~~~~~~~case~'>':~x-{}-;~break;~}{\small \par}

{\small{}~~~~~~~~~~~\}~}{\small \par}

{\small{}~~~~~~~~\$out(~n=>0~)~=~x;~}{\small \par}

{\small{}~~~~~~~~\$out(~n=>1~)~=~y;~\}~}{\small \par}

{\small{}~~~~~\%\}}{\small \par}

{\small{}\%\}}{\small \par}
\end{lyxcode}
{\small{}~~~~~free(buf);~}{\small \par}

{\small{}EOHPP~}{\small \par}

{\small{});}{\small \par}

{\small{}EOF~}{\small \par}
\end{lyxcode}
This is the end of the inline PP block, and Perl code continues below.
There's just a simple print statement, to show that we can pick up
the thread. The print executes \emph{after} Inline finishes compiling
\texttt{\small{}PDL::hpp}.
\begin{lyxcode}
{\small{}print~``Compiled~hilbert!\textbackslash{}n'';}{\small \par}
\end{lyxcode}

\subsection{\label{sub:A-standalone-PP}A standalone PP Module}

If you are coding a complete module that uses PP functions extensively,
you can place the entire module - PP, C, and Perl - in a single file
that will generate \texttt{\small{}.xs} and \texttt{\small{}.pm} files
when built. Most of PDL itself is implemented this way. There are
two important things that aren't obvious from using PP inline: (1)
you need to invoke \texttt{\small{}PDL::PP} directly to process the
source code and generate the files; and (2) most authors choose to
merge the Perl and PP segments of their module into a single source
file called ``\emph{module}\texttt{\small{}.pd}''. If you use the
standard \texttt{\small{}ExtUtils::MakeMaker }or \texttt{\small{}Module::Build},
then there are straightforward recipes to use, with routines that
are included with PDL in the module \texttt{\small{}PDL::Core::Dev}.

The various intermediate and compiled files get put into the current
working directory when the PP calls are made. If you're using \texttt{\small{}ExtUtils::MakeMaker},
that is the same directory as your \texttt{\small{}.pd} file. After
compiling your module, you'll find a \texttt{\small{}.xs}, a \texttt{\small{}.c},
a \texttt{\small{}.pm}, and a \texttt{\small{}.so} file in the same
directory. The human readable files (\texttt{\small{}.xs}, \texttt{\small{}.c},
and \texttt{\small{}.pm}) all get warning comments at the top of the
file, pointing out that they were generated from your .pd file and
should not generally be edited.

This section is a demonstration of how to implement a simple object,
\texttt{\small{}PPDemo::Hello}, as a standalone module that uses Perl,
C, and PP code. All the elements are declared inside one file, \texttt{\small{}hello.pd}.

\texttt{\small{}PPDemo::Hello} is a bit silly -- it keeps track of
the number of people in a large guest house, indexed by day of arrival
or departure and ID of their group. It tracks arrival and departure
of people in events, and integrates those events to arrive at the
number of dinner guests each day. The information is kept in a 3-column
PDL that contains (day-of-event, party-id, number of people). The
events are stored in rows. To avoid too much overhead for shuffling
memory, the PDL is allocated 100 rows at a time, and a separate counter
in the object keeps track of how many rows are used.

A \texttt{\small{}PPDemo::Hello} object doesn't have a lot of extra
information in it, so it also demonstrates PDL's autohash extensions:
the object is stored in a hash ref, and the database is in a key called
``\texttt{\small{}PDL}'' -- so it can also be manipulated directly
as a PDL object by all the usual techniques. To make that happen,
we just add ``\texttt{\small{}PDL}'' to \texttt{\small{}@PPDemo::Hello::ISA}.

If you use \texttt{\small{}ExtUtils::MakeMaker} (or \texttt{\small{}Module::Build}),
you only need two files -- a \texttt{\small{}Makefile.PL} (or \texttt{\small{}Build.PL})
and a single ``.pd'' file that invokes PP to generate the program
files.

\subsubsection{Makefile.PL}

Within the context of a MakeMaker module, you can use a simple Makefile.PL
to compile your code, drawing functionality from the \texttt{\small{}PDL::Core::Dev}
module. Here's a sample:
\begin{lyxcode}
{\small{}\#~Sample~Makefile.PL~for~a~PP-based~module}{\small \par}

{\small{}use~ExtUtils::MakeMaker;}{\small \par}

{\small{}use~PDL::Core::Dev;}{\small \par}

{\small{}PDL::Core::Dev->import();}{\small \par}

{\small{}my~\$package~=~{[}}{\small \par}
\begin{lyxcode}
{\small{}``hello.pd'',~~~~~~~\#~File~name~containing~your~module's~PP~declarations}{\small \par}

{\small{}``PPDemo\_Hello'',~~~\#~Preferred~name~for~your~module's~intermediate~files}{\small \par}

{\small{}``PPDemo::Hello''~~~\#~Fully~qualified~name~for~the~module~you're~declaring}{\small \par}

{\small{}{]};}{\small \par}
\end{lyxcode}
{\small{}my~\%hash~=~pdlpp\_stdargs(\$package);}{\small \par}

{\small{}\#~(Modify~\%hash~as~you~wish~here~-~adding~metadata,~author~name,~etc.)}{\small \par}

{\small{}WriteMakefile(~\%hash~);}{\small \par}

{\small{}sub~MY::postamble~\{~pdlpp\_postamble(\$package)~\};}{\small \par}
\end{lyxcode}
The \texttt{\small{}pdlpp\_stdargs} call loads \texttt{\small{}\%hash}
with the appropriate values to set up compilation -- it loads \texttt{\small{}\%hash}
with all the parameters needed to make MakeMaker compile your \texttt{\small{}.pd}
file into \texttt{\small{}.pm} and \texttt{\small{}.xs} files, to
compile that \texttt{\small{}.xs} file into appropriate dynamic libraries
to which Perl can link, and to put the compiled files in their correct
places in the \texttt{\small{}blib} hierarchy. Your \texttt{\small{}.pd}
file, as with any other Perl file, may or may not actually declare
a complete Perl module. For example, in the PDL distribution itself
many \texttt{\small{}.pd} files create additional functionality directly
in the \texttt{\small{}PDL} package rather than declaring their own
package.

\subsubsection{hello.pd }

The single file you use for your PP module is a ``.pd'' file. It
is a perl program that calls PP functions to emit code, so it starts
a lot like any other Perl script. The very first call is to \texttt{\small{}pp\_bless},
to make sure PP puts compiled objects into the package we want. (The
default is ``\texttt{\small{}PDL}'', but if you just dump all your
methods in there you lose the benefits of a hierarchical name space).
We also call \texttt{\small{}pp\_add\_isa} to make sure we can manipulate
our object with PDL calls. You probably don't need that for your object,
unless you want to be able to manipulate it directly as a PDL. (All
PP operators, including the main PDL API, can accept a blessed hash
ref containing a piddle in the field named ``\texttt{\small{}PDL}'',
instead of an ordinary piddle.)
\begin{lyxcode}
{\small{}\#!/usr/bin/perl~~\#~(Or~whatever~for~your~system)}{\small \par}

{\small{}\#~This~is~hello.pd~-~a~demo~file~for~a~simple~PP-based~module.}{\small \par}

use~PDL;

{\small{}pp\_bless('PPDemo::Hello');~\#~equivalent~of~``package''~at~top~of~a~.pm}{\small \par}

{\small{}pp\_add\_isa('PDL');~~~~~~~~~\#~hook~in~the~PDL~hash-extension~stuff}{\small \par}
\end{lyxcode}
The next bit is also common -- we want to stick some POD and ``\texttt{\small{}use
strict}'' into the top of our module. So we use \texttt{\small{}pp\_addpm}
to place those things into the .pm file that PP will build.
\begin{lyxcode}
{\small{}pp\_addpm(~\{At=>Top\},~<\textcompwordmark{}<'EOPM');~\#~place~following~perl~code~at~the~top~of~the~file}{\small \par}

{\small{}~}{\small \par}

{\small{}use~strict;}{\small \par}

{\small{}~}{\small \par}

{\small{}=head1~NAME}{\small \par}

{\small{}~}{\small \par}

{\small{}Hello~-~sample~perl~+~PP~object}{\small \par}

{\small{}~}{\small \par}

{\small{}=head1~DESCRIPTION}{\small \par}

{\small{}~}{\small \par}

{\small{}Hello~keeps~track~of~long-term~guests~over~a~span~of~days,~counted~by~}{\small \par}

{\small{}day~from~an~epoch.~~You~tell~it~when~guests~enter~and~leave.~~Methods}{\small \par}

{\small{}retrieve~information~about~how~full~the~house~is,~and~about~who~is~the~}{\small \par}

{\small{}most~frequent~visitor.}{\small \par}

{\small{}~}{\small \par}

{\small{}Methods~are~``new'',~``event'',~``tomorrow'',~and~``dinner\_count''.}{\small \par}

{\small{}~}{\small \par}

{\small{}=cut}{\small \par}

{\small{}~}{\small \par}

{\small{}\#\#\#~Do~this~if~you~want~your~POD~loaded~ino~the~PDL~shell's~help~database~on-the-fly.}{\small \par}

{\small{}\$PDL::onlinedoc->scan(\_\_FILE\_\_)~if(\$PDL::onlinedoc);}{\small \par}

~

{\small{}EOPM}{\small \par}
\end{lyxcode}
So much for Perl header material. We can also add some C header material.
Here, it's just a sample comment and a helper C function to print
the day of the week. The helper function uses \texttt{\small{}pdl\_malloc},
which is a fire-and-forget version of \texttt{\small{}malloc} that
produces scratch areas on the fly. \texttt{\small{}pdl\_malloc} uses
the Perl garbage collector mechanism: its returned pointer is the
data field of an anonymous mortal SV, which is cleaned up automagically
by Perl whenever you exit the enclosing scope. The C code will get
placed near the top of the generated XS file, just above the various
function declarations generated by PP itself.
\begin{lyxcode}
{\small{}pp\_addhdr(~<\textcompwordmark{}<'EOC');}{\small \par}

{\small{}~}{\small \par}

{\small{}PDL\_COMMENT(``~This~is~C.~~Both~pdl.h~and~pdlcore.h~have~been~included.~~(This~``);}{\small \par}

{\small{}PDL\_COMMENT(``~style~is~preferred~by~some~over~/{*}~{*}/,~which~doesn't~nest.)``);}{\small \par}

{\small{}~}{\small \par}

{\small{}\#include~<stdio.h>}{\small \par}

{\small{}PDL\_COMMENT(``~Return~the~abbreviated~day~of~the~week~in~a~short~scratch~string~``);}{\small \par}

{\small{}static~char~{*}dow\_data=''Sun\textbackslash{}0Mon\textbackslash{}0Tue\textbackslash{}0Wed\textbackslash{}0Tu\textbackslash{}0Fri\textbackslash{}0Sat\textbackslash{}0'';}{\small \par}

{\small{}char~{*}dow(~unsigned~int~n~)~\{}{\small \par}

{\small{}~~unsigned~int~dex~=~n~\%~7;}{\small \par}

{\small{}~~char~{*}out~=~pdl\_malloc(~4~{*}~sizeof(char)~);}{\small \par}

{\small{}~~strncpy(out,~dow\_data~+~4~{*}~dex,~4);}{\small \par}

{\small{}~~return~out;}{\small \par}

{\small{}\}}{\small \par}

{\small{}~}{\small \par}

{\small{}EOC}{\small \par}
\end{lyxcode}
Most of the methods are Perl-side, so they are placed in a single
pp\_addpm call. You can break up the call if you choose to. Since
we don't really care where this material goes in the module, we don't
specify the lcoation to \texttt{\small{}pp\_addpm}. It defaults to
the ``middle'' (after the Top material and before the Bottom material).
The Top, Bottom, and Middle queues are maintained separately, so material
appears in the order of \texttt{\small{}pp\_addpm} calls for each
of those parts of the file. Notice that the methods make a lot of
use of the fact that regular PDL methods work on the object itself
automagically since it keeps its primary piddle in the hash field
named '\texttt{\small{}PDL}'. That shortcut is apparent in the stringifier,
in particular, where we use ``\texttt{\small{}slice}'' and such
on the object itself instead of the piddle that it contains.
\begin{lyxcode}
{\small{}pp\_addpm(~<\textcompwordmark{}<'EOPM'~);}{\small \par}

{\small{}~}{\small \par}

{\small{}\#\#~Simple~constructor}{\small \par}

{\small{}sub~new~\{~}{\small \par}
\begin{lyxcode}
{\small{}return~bless(~\{~day=>0,~n=>0,~PDL=>zeroes(long,~3,~100)~\},~\$\_{[}0{]}~);}{\small \par}
\end{lyxcode}
{\small{}\}}{\small \par}

{\small{}~}{\small \par}

{\small{}\#\#~Stringifier~-{}-~so~``print~\$a''~makes~something~useful}{\small \par}

{\small{}use~overload~''''''~=>~sub~\{}{\small \par}
\begin{lyxcode}
{\small{}my~\$me~=~shift;}{\small \par}

{\small{}my~\$s~=~sprintf(``PPDemo~object~-~~days:~\%d;~~events:~\%d;~~ids:~\%d'',}{\small \par}
\begin{lyxcode}
{\small{}(\$me->slice(0)->max~-~\$me->slice(0)->min~+~1),}{\small \par}

{\small{}\$me->\{n\},~\$me->dim(1),}{\small \par}

{\small{}\$me->slice({[}0,0,0{]})->qsort->uniq->nelem}{\small \par}

{\small{});}{\small \par}
\end{lyxcode}
{\small{}\$s~.=~(\$me->\{n\})~?~}{\small \par}
\begin{lyxcode}
{\small{}\$me->slice('x',{[}0,\$me->\{n\}-1{]})~:}{\small \par}

{\small{}``\textbackslash{}n{[}~Empty~{]}\textbackslash{}n'';}{\small \par}
\end{lyxcode}
{\small{}return~\$s;}{\small \par}
\end{lyxcode}
{\small{}\};}{\small \par}

{\small{}~}{\small \par}

{\small{}\#\#~event~adder~-~add~a~row~to~the~event~database~(or~several~-~it's~threadable)}{\small \par}

{\small{}sub~event~\{}{\small \par}
\begin{lyxcode}
{\small{}my~(\$me,~\$id,~\$num)~=~(\$\_{[}0{]},~pdl(\$\_{[}1{]}),~pdl(\$\_{[}2{]}));}{\small \par}

{\small{}~}{\small \par}

{\small{}\#\#~Extend~rows~if~necessary}{\small \par}

{\small{}me->\{PDL\}~=~\$me->\{PDL\}->glue(1,zeroes(3,100))}{\small \par}

{\small{}~~unless(~\$me->\{n\}~+~\$id->dim(0))~<~\$me->\{PDL\}->dim(1)~);}{\small \par}

{\small{}~}{\small \par}

{\small{}my~\$cut~=~\$me->slice('x',{[}\$me->\{n\},~\$me->\{n\}~+~\$id->dim(0)~-~1,~1{]});}{\small \par}

{\small{}\$cut->slice({[}0,0,0{]})~.=~\$me->\{day\};}{\small \par}

{\small{}\$cut->slice({[}1,1,0{]})~.=~\$id;}{\small \par}

{\small{}\$cut->slice({[}2,1,0{]})~.=~\$type;}{\small \par}

{\small{}\$me->\{n\}~+=~\$id->dim(0);}{\small \par}

{\small{}return~\$me;}{\small \par}
\end{lyxcode}
{\small{}\}}{\small \par}

{\small{}~}{\small \par}

{\small{}\#\#~Advance~day}{\small \par}

{\small{}sub~tomorrow~\{~\$\_{[}0{]}->\{day\}++;~\}}{\small \par}

~

{\small{}EOPM}{\small \par}
\end{lyxcode}
Here's the first PP definition - a simple dinner-guest counter. You
can normally get away with just a \texttt{\small{}Code} section to
the PP definition, as in the other examples -- but here we use the
\texttt{\small{}PMCode} section to declare a Perl wrapper function
that preparses the arguments. The parsing here is pretty trivial --
it amounts to just making the \texttt{\small{}\$day} field default
to -1 if not specified, since PP itself will autogenerate null PDLs
if the output parameter isn't specified. The PP declaration just acccumulates
dinner guests by event from the beginning of time until the specified
day (or the end of the data set), and returns the sum. 

If you specify a \texttt{\small{}PMCode} field, then the \texttt{\small{}Code}
field generates a function called \texttt{\small{}\_}\emph{<name>}\texttt{\small{}\_int},
instead of just \emph{<name>}. The \texttt{\small{}PMCode} section
is a Perl snippet that must declare \emph{<name>. }Warning: If you
don't make a \texttt{\small{}pp\_bless} call (as we did above), then
the \texttt{\small{}\_<}\emph{name}\texttt{\small{}>\_int} ends up
in the package \texttt{\small{}PDL}, while the Perl declaration executes
in the current package. In that case, you'd have to call \texttt{\small{}PDL::\_}\emph{<name>}\texttt{\small{}\_int}
explicitly. But it's best to call \texttt{\small{}pp\_bless} and get
everything in the right place from the start.
\begin{lyxcode}
{\small{}pp\_def(}{\small \par}
\begin{lyxcode}
{\small{}'dinner\_count',}{\small \par}

{\small{}Pars=>'in(col=3,m);~day();~{[}o{]}out()',}{\small \par}

{\small{}GenericTypes=>{[}'L'{]},}{\small \par}

{\small{}Code=><\textcompwordmark{}<'EOC',}{\small \par}
\end{lyxcode}
{\small{}\$out()~=~0;}{\small \par}

{\small{}loop(m)~\%\{}{\small \par}

{\small{}~if(~(\$day()~<~0)~||~(\$in(~col=>0~)~<=~\$day())~)}{\small \par}

{\small{}~~~~\$out()~+=~\$in(~col=>2~);}{\small \par}

{\small{}\%\}}{\small \par}

{\small{}EOC}{\small \par}
\begin{lyxcode}
{\small{}PMCode~=>~<\textcompwordmark{}<'EOPM',}{\small \par}
\end{lyxcode}
{\small{}sub~dinner\_count~\{}{\small \par}
\begin{lyxcode}
{\small{}my(\$in,~\$day,~\$out)~=~(~(\$\_{[}0{]}),~(\$\_{[}1{]}~//~-1),~(\$\_{[}2{]}~//~PDL::null()));}{\small \par}

{\small{}\_dinner\_count\_int(\$in,~\$day,~\$out);}{\small \par}

{\small{}return~\$out;}{\small \par}
\end{lyxcode}
{\small{}\}}{\small \par}

{\small{}EOPM}{\small \par}

{\small{});}{\small \par}
\end{lyxcode}
Here's the second (and last) PP definition -- this one returns a string
listing dinner attendees by day. It might be better accomplished using
Perl, but this demonstrates how to return a Perl SV from a PP routine.
There's no clean mechanism to do so, so it works by modifying an SV
in place and passing that out through the PMCode wrapper. The PMCode
wrapper generates an empty string SV and passes in a reference to
it. \texttt{\small{}dinner\_plan} accesses the referred-to SV, and
appends to the contained string. That same sort of hand-back works
for other Perl types that aren't PDLs, as well. There is an explicit
\texttt{\small{}threadloop} in the \texttt{\small{}dinner\_plan} PP
code, to avoid reallocating the char buffer on every thread iteration.
\begin{lyxcode}
~{\small{}pp\_def(~\#~Returns~a~Perl~scalar~by~handing~a~ref~into~the~compiled~code}{\small \par}
\begin{lyxcode}
{\small{}'dinner\_plan',~	~~~~~~~~}{\small \par}

{\small{}Pars=>'in(col=3,m);~day()',~	}{\small \par}

{\small{}OtherPars=>'SV~{*}rv',~	~~~}{\small \par}

{\small{}GenericTypes=>{[}'L'{]},~	}{\small \par}

{\small{}Code=>~<\textcompwordmark{}<'EOC',~}{\small \par}
\end{lyxcode}
{\small{}char~buffer{[}BUFSIZ{]};~}{\small \par}

{\small{}threadloop~\%\{}{\small \par}
\begin{lyxcode}
{\small{}int~day~=~\$day();~}{\small \par}

{\small{}int~nguests~=~0;~}{\small \par}

{\small{}PDL\_COMMENT(``~sv~is~a~Perl~scalar~in~which~we~want~to~return~a~value.~``)}{\small \par}

{\small{}PDL\_COMMENT(``~We~do~this~by~passing~in~a~ref~to~it,~and~following~the~ref~here.'')}{\small \par}

{\small{}SV~{*}sv~=~SvRV(\$COMP(rv));~~}{\small \par}

{\small{}if(day<0)~\{~~~~/{*}~Negative~input~day~means~``use~the~last~day''.~~{*}/}{\small \par}
\begin{lyxcode}
{\small{}loop(m)~\%\{~~/{*}~Accumulate~the~maximum~value~of~the~day~field~{*}/~~~~~~~~~~~~}{\small \par}
\begin{lyxcode}
{\small{}if(\$in(col=>0)~>~day)~~~~~}{\small \par}
\begin{lyxcode}
{\small{}day~=~\$in(col=>0);~~~~~~~~~~~~~}{\small \par}
\end{lyxcode}
\end{lyxcode}
{\small{}\%\}~~~~~~~~~~~~}{\small \par}
\end{lyxcode}
{\small{}\}~}{\small \par}

{\small{}loop(m)~\%\{~~/{*}~Accumulate~dinner~guests~for~this~day~{*}/}{\small \par}
\begin{lyxcode}
{\small{}if(\$in(col=>0)~<=~day)~~~~~~}{\small \par}
\begin{lyxcode}
{\small{}nguests~+=~\$in(col=>2);~}{\small \par}
\end{lyxcode}
\end{lyxcode}
{\small{}\%\}}{\small \par}

{\small{}/{*}~Append~a~summary~of~the~day~to~the~passed-in~string~{*}/}{\small \par}

{\small{}sprintf(buffer,\textquotedbl{}Day~\%d~(\%s):~plan~for~\%d\textbackslash{}n\textquotedbl{},~day,~dow(day),~nguests);~}{\small \par}

{\small{}sv\_catpv(sv,~buffer);~}{\small \par}
\end{lyxcode}
{\small{}\%\}}{\small \par}

{\small{}EOC}{\small \par}

{\small{}	PMCode~=>~<\textcompwordmark{}<'EOPM'	~}{\small \par}

{\small{}sub~dinner\_plan~\{~~~}{\small \par}
\begin{lyxcode}
{\small{}my(\$in,~\$day)~=~(~(\$\_{[}0{]}),~(\$\_{[}1{]}~//~-1)~);~~~}{\small \par}

{\small{}my~\$s~=~\textquotedbl{}\textquotedbl{};~~~~~~~~~~~~~~~~~~~~~~~~\#~Generate~an~empty~string~\$s}{\small \par}

{\small{}\_dinner\_plan\_int(\$in,~\$day,~\textbackslash{}\$s);~~\#~hand~in~a~ref~to~\$s~(to~be~modified~inplace)}{\small \par}

{\small{}return~\$s;~\}~}{\small \par}
\end{lyxcode}
{\small{}EOPM~}{\small \par}

{\small{});~}{\small \par}
\end{lyxcode}
Finally, we have to finish the output with:
\begin{lyxcode}
{\small{}pp\_done();}{\small \par}
\end{lyxcode}
The earlier declarations only modify package-global variables in the
PP code generator. \texttt{\small{}pp\_done()} does the actual work
of generating the files.

\subsubsection{Using PPDemo::Hello}

Here's a sample interaction with the perldl shell. User input is in
bold. 
\begin{lyxcode}
{\small{}pdl>~}\textbf{\small{}use~PPDemo::Hello}{\small \par}

{\small{}pdl>~}\textbf{\small{}\$a~=~new~PPDemo::Hello}{\small \par}

{\small{}pdl>~}\textbf{\small{}\$a->event(~pdl(1,6,7,8),~pdl(-1,-1,2,3)~);}{\small \par}

{\small{}pdl>~}\textbf{\small{}\$a->tomorrow;}{\small \par}

{\small{}pdl>~}\textbf{\small{}\$a->event(~1,~3~);}{\small \par}

{\small{}pdl>~}\textbf{\small{}\$a->event(~9,~1~);}{\small \par}

{\small{}pdl>~}\textbf{\small{}p~\$a}{\small \par}

{\small{}PPDemo~object~-~~days:~2;~~events:~6/100;~~ids:~2}{\small \par}

{\small{}{[}}{\small \par}

{\small{}~{[}~0~~1~-1{]}}{\small \par}

{\small{}~{[}~0~~6~-1{]}}{\small \par}

{\small{}~{[}~0~~7~~2{]}}{\small \par}

{\small{}~{[}~0~~8~~3{]}}{\small \par}

{\small{}~{[}~1~~1~~3{]}}{\small \par}

{\small{}~{[}~1~~9~~1{]}}{\small \par}

{\small{}{]}}{\small \par}

{\small{}pdl>~}\textbf{\small{}p~\$a->dinner\_plan();}{\small \par}

{\small{}Day~1~(Mon):~plan~for~7}{\small \par}

{\small{}pdl>~}\textbf{\small{}p~\$a->dinner\_plan(pdl(0,1,2));}{\small \par}

{\small{}Day~0~(Sun):~plan~for~3}{\small \par}

{\small{}Day~1~(Mon):~plan~for~7}{\small \par}

{\small{}Day~2~(Tue):~plan~for~7}{\small \par}

{\small{}pdl>~}\textbf{\small{}p~\$a->dinner\_count}{\small \par}

{\small{}7}{\small \par}

{\small{}pdl>}{\small \par}
\end{lyxcode}
The object works as expected, and the C and \texttt{\small{}Perl}
aspects merge (practically) seamlessly. In particular, note that \texttt{\small{}dinner\_plan}
threads normally - but the output is still just a Perl scalar string
that happens to contain all the days, summarized by line.

\section{Using PP for Calculations (no dataflow)}

The most basic use of PP is to produce a single output PDL that is
obtained by acting on one or more inputs. The examples in §\ref{sec:Basic-PP}-§\ref{sec:Putting-it-together:}
all happen to be simple calculation cases, producing a single output.
But you can return any value or multiplet of values you want, by specifying
multiple {[}o{]} variables in the \texttt{\small{}Pars} signature,
or by accepting refs in the \texttt{\small{}OtherPars} field. Direct
assignment, temporary variables, and Perl wrappers have all been demonstrated
in §\ref{sec:Basic-PP}-§\ref{sec:Putting-it-together:} but are described
here in §\ref{sub:Using-direct-assignment}-§\ref{sub:Conditioning-arguments-with}
in more depth. Bad value handling is straightforward and in §\ref{sub:BAD-values}.

\subsection{\label{sub:Using-direct-assignment}Using direct assignment}

The simplest and most common type of PP operator calculates a value
and stuffs it into an appropriate location in an output PDL. You design
your algorithm to act on its simplest input data type (e.g. a scalar,
a vector, a collection of vectors, or what-have-you), and write a
signature to match that data type. You declare an output variable
in the signature, assume it is allocated correctly for your input
parameter, and assign your computed value directly to the element(s)
of the output variable. That basic mode of operation is in most of
the examples above, including §\ref{sub:A-simple-PP}. 

Direct assignment doesn't necessarily have to be simple. Here is an
example of a function that accepts an NxM array and returns its average
(a scalar), its average-over-columns (an M-vector) and its average-over-rows
(an N-vector):
\begin{lyxcode}
{\small{}pp\_def('multisum',}{\small \par}

{\small{}~~~~~~~Pars=>'im(n,m);~{[}o{]}av();~{[}o{]}avc(m);~{[}o{]}avr(n);',}{\small \par}

{\small{}~~~~~~~Code~=>~<\textcompwordmark{}<'EOC',}{\small \par}
\begin{lyxcode}
{\small{}\$av()~=~0;}{\small \par}

{\small{}loop(n)~\%\{~\$avr()~=~0;~\%\}~//~initialize~avr}{\small \par}

{\small{}loop(m)~\%\{~}{\small \par}

{\small{}~~\$avc()~=~0;~~~~~~~~~~~~~//~initialize~avc~element}{\small \par}

{\small{}~~loop(n)~\%\{}{\small \par}

{\small{}~~~~\$GENERIC()~pix~=~\$im();}{\small \par}

{\small{}~~~~\$av()~~+=~pix;}{\small \par}

{\small{}~~~~\$avc()~+=~pix;}{\small \par}

{\small{}~~~~\$avr()~+=~pix;}{\small \par}

~~\%\}

~~{\small{}\$avc~/=~\$SIZE(n);~~~~~~~//~normalize~avc~element}{\small \par}

{\small{}\%\}}{\small \par}

{\small{}loop(n)~\%\{~\$avr()~/=~\$SIZE(m);~\%\}~~//~normalize~avr}{\small \par}

{\small{}\$av()~/=~\$SIZE(m)~{*}~\$SIZE(n);~~~~~~//~normalize~av}{\small \par}
\end{lyxcode}
{\small{}EOC}{\small \par}

{\small{}Doc~=>~<\textcompwordmark{}<'EOD'}{\small \par}

{\small{}=for~ref}{\small \par}

{\small{}Sample~PP~code~uses~direct~assignment~to~a~scalar~and~two~vectors.~The~input~has}{\small \par}

two~active~dims~named~'n'~and~'m'.~~We~loop~over~both,~accumulating~a~

(scalar)~average~of~the~whole~input~matrix,~a~(m-vector)~average~across~

each~row,~and~a~(n-vector)~average~across~each~column.

~

=cut

EOD

{\small{});}{\small \par}
\end{lyxcode}
Here, we accumulate sums three different ways and then normalize them
appropriately. Note that the \texttt{\small{}m} and \texttt{\small{}n}
values are automatically handled right by the index macros inside
the nested loops. The \texttt{\small{}\$av()} macro gets indexed with
nothing, the \texttt{\small{}\$avc()} macro gets indexed with the
implicit value of \texttt{\small{}m}, and the \texttt{\small{}\$avr()
}macro gets indexed with the implicit value of \texttt{\small{}n}.

\subsection{\label{sub:Temporary-variables}Temporary Variables to Ease the Pain}

Nearly any nontrivial calculation requires holding intermediate values
in local variables. For scalar quantities this is simple -- you can
declare a C variable in the Code section of your PP declaration. If
you want to shave a few clock cycles off your calculation, you can
even declare the C variable outside an explicit \texttt{\small{}threadloop
\%\{ \%\}} block, to prevent it being pushed onto the stack (and then
popped off again) every time the thread loop executes. For vector
quantities it is not so simple - C requires you to track the size
explicitly, and generally to allocate the memory too. 

Rather than explicitly declaring an array that is just the right size,
you can declare a temporary variable in the signature of your PP function,
and use it just like a passed-in PDL. The access macros will work
correctly, and the temporary variable is typed just exactly like other
arguments (e.g. \texttt{\small{}\$GENERIC()} or other type as declared).
The temporary variable is declared just large enough to hold one instance
of its declared active dims -- any thread dimensions are ignored,
so you don't allocate extra memory you don't need. (Warning: Temporary
variables are currently not threadsafe in PDL 2.007. Threadsafe temporaries
are a planned feature for PDL 3.)

Here's an example that uses a temporary variable to stash the denominator
and discriminant of the quadratic equation. These could have been
declared as scalar variables - putting them in a temporary PDL demonstrates
autodeclaration of a vector quantity. 
\begin{lyxcode}
{\small{}no~PDL::NiceSlice;}{\small \par}

{\small{}{*}solve\_quad~=~\textbackslash{}\&PDL::solve\_quad;}{\small \par}

{\small{}use~Inline~Pdlpp~=>~<\textcompwordmark{}<'EOF';}{\small \par}

{\small{}pp\_def('solve\_quad',}{\small \par}
\begin{lyxcode}
{\small{}Pars=>'coeffs(n=3);~{[}o{]}sols(s=2);~{[}t{]}parts(s);',}{\small \par}

{\small{}GenericTypes=>{[}F,D,S{]},}{\small \par}

{\small{}Code~=>~<\textcompwordmark{}<'EOC',}{\small \par}
\end{lyxcode}
{\small{}/{*}~Stash~the~denominator~and~discriminant~{*}/}{\small \par}

{\small{}\$parts(s=>0)~=~1.0~/~(\$coeffs(n=>2)~{*}~2);}{\small \par}

{\small{}\$parts(s=>1)~=~\$coeffs(n=>1)~{*}{*}~2~-~4~{*}~\$coeffs(n=>2)~{*}~\$coeffs(n=>0);}{\small \par}

{\small{}~}{\small \par}

{\small{}if(\$parts(s=>1)~>=~0)~\{}{\small \par}
\begin{lyxcode}
{\small{}/{*}~One~or~more~solutions~exist.~~~Put~them~in~the~output.~{*}/}{\small \par}

{\small{}\$parts(s=>1)~=~sqrt(\$parts(s=>1));}{\small \par}

{\small{}\$sols(s=>0)~=~\$parts(s=>0)~{*}~(~-\$coeffs(n=>1)~-~\$parts(s=>1)~);}{\small \par}

{\small{}\$sols(s=>1)~=~\$parts(s=>0)~{*}~(~-\$coeffs(n=>1)~+~\$parts(s=>1)~);}{\small \par}
\end{lyxcode}
{\small{}\}~else~\{}{\small \par}
\begin{lyxcode}
{\small{}/{*}~No~real~solutions~exist.~~Set~answer~to~NaN.~~notice~the~type~selection~macro!~{*}/}{\small \par}

{\small{}\$sols(s=>0)~=~\$sols(s=>1)~=~\$TFDS(~union\_nan\_float.f,~union\_nan\_double.d,~-32768~);}{\small \par}
\end{lyxcode}
{\small{}\}}{\small \par}

{\small{}EOC}{\small \par}

{\small{});}{\small \par}

{\small{}EOF}{\small \par}
\end{lyxcode}
The '\texttt{\small{}solve\_quad}' example solves a quadratic equation
analytically. The discriminant and denominator of the quadratic formula
are used more than once in valid solutions, so they get stashed in
the temporary variable \texttt{\small{}parts}, which is declared as
a 2-vector. In the event that the discriminant is negative, no real
solutions exist, and the routine returns the appropriately typed NaN.
In cases where a specific action is needed for a specific type, you
can use the \texttt{\small{}\$T} macro -- you list the \texttt{\small{}\$T}
followed by a collection of one-letter type identifiers. The code
switches between the strings in the argument list, based on the current
value of \texttt{\small{}\$GENERIC()} -- so the floating-point version
of the code (``\texttt{\small{}F}'') gets \texttt{\small{}``union\_nan\_float.f}''
and the double-precision version (``\texttt{\small{}D}'') gets ``\texttt{\small{}union\_nan\_double.d}''.
Each of those expressions contains an appropriate NaN value for that
data type. Short types ('\texttt{\small{}S}') can't support NaN, so
we supply -32768 instead. This would also be a good place to use BAD
values instead of NaN (see §\ref{sub:BAD-values}).

\subsection{\label{sub:Conditioning-arguments-with}Conditioning arguments with
Perl}

Surprisingly often you might want to change the default behavior for
a PP operator, or prepare or curry arguments in some way, before entering
the actual algorithm. For non-dataflow PP operators, there's a mechanism
to do that. If you specify a field called \texttt{\small{}PMCode}
in your PP declaration, the compiled stuff from the \texttt{\small{}Code}
field will go into \texttt{\small{}\_}\emph{subname}\texttt{\small{}\_int
}instead of just \emph{subname}. Your \texttt{\small{}PMCode} field
can then contain some Perl that declares \emph{subname} and (optionally)
springboards into \texttt{\small{}\_}\emph{subname}\texttt{\small{}\_int}.
Here's an example: a routine calle\texttt{\small{}d minismooth} that
accepts a 2-D PDL image and returns its input, smoothed by a quasi-minimum
operator. We use a temporary variable, too, to demonstrate how to
use it from a Perl-wrapped call.

\texttt{\small{}minismooth} is useful, for example, for finding a
smooth background image in the presence of spiky foreground objects
such as a starfield: the stars are eliminated in the local minimum-smoothing
process, leaving locally ``typical'' values for the background.

The arguments to \texttt{\small{}minismooth} get conditioned by the
PMCode, so that it can be called more flexibly than a naked PP operator
can.

The \texttt{\small{}minismooth} algorithm itself is straightforward.
There are two active dims -- the height and width of the image. The
code loops over all pixels in the image, and for each one it loops
over the entire neighborhood, accumulating the \emph{n} lowest values
in the image. The output pixel at the current location gets the \emph{n}th
lowest value from the neighborhood.

(Note that in versions of PDL up to at least 2.009, this type of declaration
(using \texttt{\small{}PMCode}) only works inside a standalone module
definition. A long-standing wart in \texttt{\small{}Inline::Pdlpp}
prevents the \texttt{\small{}PMCode} from being interpreted.)
\begin{lyxcode}
{\small{}pp\_def('minismooth',}{\small \par}
\begin{lyxcode}
{\small{}Pars=>'im(n,m);~{[}o{]}sm(n,m);~{[}t{]}list(n);',}{\small \par}

{\small{}OtherPars=>'long~size;~long~nth;',}{\small \par}

{\small{}PMCode~=>~<\textcompwordmark{}<'EOPMC',}{\small \par}
\end{lyxcode}
{\small{}=head2~minismooth}{\small \par}

{\small{}~}{\small \par}

{\small{}=for~usage}{\small \par}

{\small{}~}{\small \par}
\begin{lyxcode}
{\small{}\$sm~=~minismooth(~\$im,~\$size,~\$nth~);}{\small \par}
\end{lyxcode}
{\small{}~}{\small \par}

{\small{}=for~ref}{\small \par}

{\small{}~}{\small \par}

{\small{}C<minismooth>~smooths~an~image~by~finding~the~minimum~(or~near-minimum)~over~}{\small \par}

{\small{}the~C<\$size>xC<\$size>~block~of~pixels~around~each~source~pixel.~~If~you~specify~}{\small \par}

{\small{}C<\$nth>,~then~the~nth~smallest~pixel~in~each~neighborhood~is~used,~instead~of~the}{\small \par}

{\small{}absolute~smallest.}{\small \par}

{\small{}~}{\small \par}

{\small{}=cut}{\small \par}

{\small{}sub~PDL::minismooth~\{}{\small \par}
\begin{lyxcode}
{\small{}my~\$im~=~shift;}{\small \par}

{\small{}die~\textquotedbl{}minismooth~requires~a~PDL!\textquotedbl{}~unless(~UNIVERSAL::isa(\$im,'PDL'));}{\small \par}

{\small{}my~\$size~=~shift~//~21;}{\small \par}

{\small{}my~\$nth~=~shift~//~1;}{\small \par}

{\small{}\$nth~=~1~if(\$nth~~<~1);}{\small \par}

{\small{}\$size=1~~if(\$size~<~1);}{\small \par}

{\small{}my~\$sm~=~PDL->null;}{\small \par}

{\small{}PDL::\_minismooth\_int(\$im,~\$sm,~PDL->null,~\$size,~\$nth);}{\small \par}

{\small{}return~\$sm;}{\small \par}
\end{lyxcode}
{\small{}\}}{\small \par}

{\small{}EOPMC}{\small \par}
\begin{lyxcode}
{\small{}Code~=>~<\textcompwordmark{}<'EOC',}{\small \par}
\end{lyxcode}
{\small{}\{}{\small \par}
\begin{lyxcode}
{\small{}long~i,j,k,l,ii,jj;~//~declare~index~variables}{\small \par}

{\small{}PDL\_Indx~=~(\$COMP(size)-1)/2;}{\small \par}

{\small{}long~nn,~mm;}{\small \par}

{\small{}long~n\_kept;}{\small \par}

{\small{}\$GENERIC()~current;}{\small \par}

{\small{}for(i=0;~i<\$SIZE(m);~i++)~\{~}{\small \par}
\begin{lyxcode}
{\small{}for(j=0;~j<\$SIZE(n);~j++)~\{}{\small \par}
\begin{lyxcode}
{\small{}n\_kept~=~0;}{\small \par}

{\small{}PDL\_COMMENT(\textquotedbl{}Loop~over~the~neighborhood~around~the~current~pixel\textquotedbl{});}{\small \par}

{\small{}for(k=-sz;~k<=sz;~k++)~\{}{\small \par}
\begin{lyxcode}
{\small{}for(l=-sz;~l<=sz;~l++)~\{}{\small \par}
\begin{lyxcode}
{\small{}nn~=~j+l;~}{\small \par}

{\small{}if(nn<0)~\{~nn=0;~\}~else~if(nn>=\$SIZE(n))~\{~nn=\$SIZE(n)-1;~\}}{\small \par}

{\small{}mm~=~i+k;}{\small \par}

{\small{}if(mm<0)~\{~mm=0;~\}~else~if(mm>=\$SIZE(m))~\{~mm=\$SIZE(m)-1;~\}}{\small \par}

{\small{}current~=~\$im(n=>nn,~m=>mm);}{\small \par}

{\small{}if(~(n\_kept~<~\$COMP(nth))~||~current~<~\$list(n=>0)~)~\{}{\small \par}

{\small{}~~PDL\_COMMENT(\textquotedbl{}The~current~number~fits~on~the~minimum-values~list\textquotedbl{});}{\small \par}

{\small{}~~PDL\_COMMENT(\textquotedbl{}Use~insertion~sort~to~keep~the~list~sorted\textquotedbl{});~~}{\small \par}

{\small{}~~for(ii=0;~ii<n\_kept~\&\&~\$list(n=>ii)~>=~current;~ii++)~\{~;~\}}{\small \par}

{\small{}~~if(n\_kept~<~\$COMP(nth))~\{}{\small \par}

{\small{}~~~~PDL\_COMMENT(\textquotedbl{}Few~enough~elements~-~put~'em~on~the~back\textquotedbl{});}{\small \par}

{\small{}~~~~for(jj=n\_kept;~jj>ii;~jj-{}-)~\{~\$list(n=>jj)~=~\$list(n=>jj-1);~\}}{\small \par}

{\small{}~~~~n\_kept++;}{\small \par}

{\small{}~~\}~else~\{}{\small \par}

{\small{}~~~~PDL\_COMMENT(\textquotedbl{}List~is~full~-~bump~something~off~the~front\textquotedbl{});}{\small \par}

{\small{}~~~~ii-{}-;}{\small \par}

{\small{}~~~~for(jj=0;~jj<ii;~jj++)~\{~\$list(n=>jj)~=~\$list(n=>jj+1);~\}}{\small \par}

{\small{}~~\}}{\small \par}

{\small{}~~\$list(n=>ii)~=~current;}{\small \par}

{\small{}\}~//~end~of~list~manipulation}{\small \par}
\end{lyxcode}
{\small{}\}~//~end~of~l~neighborhood~loop}{\small \par}
\end{lyxcode}
{\small{}\}~//~end~of~k~neighborhood~loop}{\small \par}

{\small{}~}{\small \par}

{\small{}\$sm(n=>j,~m=>i)~=~\$list(n=>0);}{\small \par}
\end{lyxcode}
{\small{}\}~//~end~of~j~pixel~loop}{\small \par}
\end{lyxcode}
{\small{}\}~//~end~of~i~pixel~loop}{\small \par}
\end{lyxcode}
{\small{}\}~//~end~of~PP~code}{\small \par}

{\small{}EOC}{\small \par}

{\small{});}{\small \par}

\end{lyxcode}

\subsection{\label{sub:BAD-values}BAD values}

PDL's BAD value system uses in-band special values to mark bad or
missing points in your data. Because bad values cost extra computing
time (every value handled by a bad-aware PP operator has to be checked
against the bad value by the operator), PP lets you specify two different
versions of your code - one that handles bad values and one that doesn't.
The two parallel copies of the code should implement exactly the same
function, but one with bad value handling and one without.

Each PDL carries a status flag (the badflag) that indicates whether
there \emph{might} be bad values in the PDL: if the badflag is 0,
then the PDL is known to have no bad values and the normal Code gets
used. If it is 1, then the special BadCode gets used.

You can create a BAD-aware PP operator by setting the \texttt{\small{}HandleBad}
field in the hash you pass to \texttt{\small{}pp\_def}. If HandleBad
is set to a true value, then PP looks for both a \texttt{\small{}Code}
section and a separate \texttt{\small{}BadCode} section. The \texttt{\small{}BadCode}
gets used on PDLs with a true BadFlag and the \texttt{\small{}Code}
gets used on everything else.

If you do not do anything about BAD values, then your code will ignore
them by default -- but will pass its parameters' BadFlag status on
to any output PDLs. If BAD values will cause your code trouble, you
can set the \texttt{\small{}HandleBad} field (in the pp\_def argument
hash) to 0. That will cause PP to throw an error if your code gets
a parameter with the BadFlag set.

Inside either your \texttt{\small{}BadCode} or your \texttt{\small{}Code}
section, the following macros will come in handy:
\begin{description}
\item [{\texttt{\small{}\$ISBAD(}\textit{\small{}var}\texttt{\small{})}}] This
boolean macro returns TRUE if the contained value in var is bad. The
\emph{var} is a regular PDL access macro just like you'd use to access
any of the PDLs in the signature, except that it doesn't have a leading
\texttt{\small{}\$} -- as in ``\texttt{\small{}\$ISBAD(a(n=>2))}''.
\item [{\texttt{\small{}\$ISGOOD(}\textit{\small{}var}\texttt{\small{})}}] This
is the counterpart to \$ISBAD, and returns a flag indicating goodness.
\item [{\texttt{\small{}\$SETBAD(}\textit{\small{}var}\texttt{\small{})}}] This
assigns the BAD value to the corresponding PDL argument. The \emph{var
}uses the same format as \texttt{\small{}\$ISGOOD} and \texttt{\small{}\$ISBAD}.
Unlike those two, it makes sense to use in a \texttt{\small{}Code}
section (since you may have to set the first BAD value in a particular
PDL). In the \texttt{\small{}Code} section, if you set a value to
BAD you should also set the PDL's badflag with the \texttt{\small{}\$PDLSTATESETBAD()}
macro, below
\item [{\texttt{\small{}\$ISBADVAR(}\textit{\small{}c\_var,~pdl}\texttt{\small{})}}] This
works like \texttt{\small{}\$ISBAD}, except that the first argument
is a cached value of the PDL in a \texttt{\small{}\$GENERIC()} type
C variable. The second argument is the name of the PDL it came from
(which must be one of the Pars in the signature).
\item [{\texttt{\small{}\$ISGOODVAR(}\textit{\small{}c\_var,~pdl}\texttt{\small{})}}] This
works like \texttt{\small{}\$ISGOOD}, except that the first argument
is a cached value of the PDL in a \texttt{\small{}\$GENERIC()} type
C variable. The second argument is the name of the PDL it came from
(which must be one of the Pars in the signature).
\item [{\texttt{\small{}\$SETBADVAR(}\textit{\small{}c\_var,~pdl}\texttt{\small{})}}] This
works like \texttt{\small{}\$SETBAD}, except that the first argument
is a cached value of the PDL in a \texttt{\small{}\$GENERIC()} type
C variable. Just like \texttt{\small{}\$SETBAD}, you should also set
the state of the PDL to BAD using \texttt{\small{}\$PDLSTATESETBAD()}
if you use this inside a \texttt{\small{}Code} block.
\item [{\texttt{\small{}\$PDLSTATEISBAD(}\emph{pdl}\texttt{\small{})}}] This
accesses the \emph{pdl}'s badflag, which indicates whether the PDL
\emph{may} have BAD values in it (and therefore requires bad-value
checking). You don't normally have to read the badflag explicitly,
since PP does that for you (and shunts execution into your \texttt{\small{}Code}
or \texttt{\small{}BadCode} segment accordingly). The \emph{pdl} argument
is the name of a parameter in the \texttt{\small{}Pars} field of \texttt{\small{}pp\_def}. 
\item [{\texttt{\small{}\$PDLSTATEISGOOD(}\emph{pdl}\texttt{\small{})}}] Counterpart
to \texttt{\small{}\$PDLSTATEISBAD}.
\item [{\texttt{\small{}\$PDLSTATESETBAD(}\emph{pdl}\texttt{\small{})}}] This
sets the \emph{pdl}'s badflag. You need to do this explicitly if you
set a value to BAD inside your \texttt{\small{}Code} block. The \emph{pdl}
argument is the name of a parameter in the \texttt{\small{}Pars} field
of \texttt{\small{}pp\_def}.
\item [{\texttt{\small{}\$PDLSTATESETGOOD(}\emph{pdl}\texttt{\small{})}}] This
clears the \emph{pdl}'s badflag. You shouldn't \emph{need} to do this,
but you might like to if you are in a position to \emph{know} that
none of the elements of the \emph{pdl} are BAD.
\end{description}

\subsubsection{\label{sub:BAD-value-handling:-1}BAD value handling: a simple case}

Here's a simple example of BAD value handling. The \texttt{\small{}countbad}
operator collapses a vector PDL by counting the number of BAD values
in it.
\begin{lyxcode}
{\small{}pp\_def('countbad',}{\small \par}
\begin{lyxcode}
{\small{}Pars=>'in(n);~{[}o{]}out();',}{\small \par}

{\small{}HandleBad~=>~1,}{\small \par}

{\small{}Code~=>~<\textcompwordmark{}<~'EOC',}{\small \par}
\end{lyxcode}
{\small{}\$out()~=~0;~}{\small \par}

{\small{}EOC}{\small \par}
\begin{lyxcode}
{\small{}BadCode~=>~<\textcompwordmark{}<'EOBC'}{\small \par}
\end{lyxcode}
{\small{}long~bc~=~0;}{\small \par}

{\small{}loop(n)~\%\{~}{\small \par}
\begin{lyxcode}
{\small{}if(~\$ISBAD(n)~)}{\small \par}
\begin{lyxcode}
{\small{}bc++;}{\small \par}
\end{lyxcode}
\end{lyxcode}
{\small{}\%\}}{\small \par}

{\small{}\$out()~=~bc;}{\small \par}

{\small{}EOBC}{\small \par}
\begin{lyxcode}
{\small{});}{\small \par}
\end{lyxcode}
\end{lyxcode}
The code is pretty straightforward. If the input PDL has its \texttt{\small{}badflag}
clear, i.e. it has no BAD values, then the Code section gets used.
It just returns 0. If the input PDL has its \texttt{\small{}badflag}
set, the BadCode gets used. It loops over the input values, counting
the bad values, and returns the count. The count is cached in a local
variable, \texttt{\small{}bc}, to avoid the overhead of making the
\texttt{\small{}\$out()} macro call for each iteration of the loop.

\subsubsection{\label{sub:BAD-value-handling:}BAD value handling: marking values
BAD.}

Here's an example of how to mark values BAD. The \texttt{\small{}recip}
operator implements reciprocals, with checking for the 1/0 case. There
are two copies of the reciprocal code -- the \texttt{\small{}Code}
block and the \texttt{\small{}BadCode} block. In the \texttt{\small{}Code}
block, no bad values are expected in the input, but we may have to
generate some for the output. In the \texttt{\small{}BadCode} block,
the input may contain bad values, but the badflag of the output is
automatically set before our code block gets called. See also the
example in Section \ref{sub:Adjusting-output-dimensions}.
\begin{lyxcode}
{\small{}pp\_def('recip',}{\small \par}
\begin{lyxcode}
{\small{}Pars=>'in(n);~{[}o{]}out();',}{\small \par}

{\small{}HandleBad~=>~1,}{\small \par}

{\small{}GenericTypes=>{[}'F','D'{]},~\#~only~makes~sense~for~floating-point~types}{\small \par}

{\small{}Code~=>~<\textcompwordmark{}<~'EOC',}{\small \par}
\end{lyxcode}
{\small{}if~(\$in()==0)~\{}{\small \par}
\begin{lyxcode}
{\small{}\$PDLSTATESETBAD(out());~~}{\small \par}

{\small{}\$SETBAD(out());}{\small \par}
\end{lyxcode}
{\small{}\}~else~\{}{\small \par}
\begin{lyxcode}
{\small{}\$out()~=~1.0/\$in();}{\small \par}
\end{lyxcode}
{\small{}\}}{\small \par}

{\small{}EOC}{\small \par}
\begin{lyxcode}
{\small{}BadCode~=>~<\textcompwordmark{}<'EOBC'}{\small \par}
\end{lyxcode}
{\small{}if~(~\$in()==0~||~\$ISBAD(in())~)~\{~}{\small \par}
\begin{lyxcode}
{\small{}\$SETBAD(out());}{\small \par}
\end{lyxcode}
{\small{}\}~else~\{}{\small \par}
\begin{lyxcode}
{\small{}\$out()~=~1.0/\$in();}{\small \par}
\end{lyxcode}
{\small{}\}}{\small \par}

{\small{}EOBC}{\small \par}
\begin{lyxcode}
{\small{});}{\small \par}
\end{lyxcode}
\end{lyxcode}

\subsection{\label{sub:Adjusting-output-dimensions}Adjusting output dimensions
(RedoDimsCode)}

Sometimes you don't know the dimensionality of an output variable
at compile time -- that is to say, the \texttt{\small{}\$SIZE()} of
one of the named dims in a signature argument (in the \texttt{\small{}Pars}
section of \texttt{\small{}pp\_def}) may require some computation
to figure out. PP can handle that, in a limited way. You accomplish
that by passing in the \texttt{\small{}RedoDimsCode} field into \texttt{\small{}pp\_def}.
Your \texttt{\small{}RedoDimsCode} is called during setup for the
computation (before the actual computation in the \texttt{\small{}Code}
and/or \texttt{\small{}BadCode} sections), and you can treat the \texttt{\small{}\$SIZE()}
macro as an lvalue (i.e. assign to it). 

Many of the same other macros that are available in the \texttt{\small{}Code}
and \texttt{\small{}BadCode} sections are available to you in the
\texttt{\small{}RedoDimsCode} section. The ones that are unavailable
\texttt{\small{}loop(}\emph{foo}\texttt{\small{})} and \texttt{\small{}threadloop
}loop-management macros, and the data-access \texttt{\small{}\$}\emph{var}\texttt{\small{}(}\emph{dim}=>\emph{expr...}\texttt{\small{})
macro.}{\small \par}

This lets you adjust the size of your output and/or temporary parameters
before carrying out the computation. Any thread dimensions from the
overall calculation get added onto the end of the \texttt{\small{}{[}o{]}}
output variables' dimlists automatically, after you set the \texttt{\small{}\$SIZE()}.
(That doesn't happen for \texttt{\small{}{[}t{]}} temporary variables,
since each temporary variable only gets its active dims -- so no thread
dims get tacked on the end.)

The adjustment is a bit limited, though -- you can only compute the
new size based on the information available \emph{before} the computation
takes place, since no actual computation has happened before the \texttt{\small{}RedoDimsCode}
gets called. That works easily for operations where you know the output
dimensions from the input parameters or input dimensions, and it's
common to see a one-line \texttt{\small{}RedoDimsCode} section in
those cases. The only catch is that the \texttt{\small{}\$SIZE()}
macro is write-only at that stage, so you have to read any dimensions
directly from the C structure of the PDL itself.

Operations that output sets (such as \texttt{\small{}which()}) require
actually walking through the data to figure out the output dimensions.
That can be tricky, since the threading loop constructs aren't available
in \texttt{\small{}RedoDimsCode}. The \texttt{\small{}which()} operator
and its relatives use a \texttt{\small{}PMCode} section to flatten
the input PDL, and treat it as a single array which can be looped
over explicitly with a C \texttt{\small{}for} loop. That trick can
work for \emph{n }dimensional inputs as well, provided that you know
\emph{n }when you are writing the code. 

Here is an example of a simple \texttt{\small{}RedoDimsCode} that
uses just the input dimensions to change the output. For more complex
examples (including the \texttt{\small{}which()} trick and more powerful
techniques such as are used by \texttt{\small{}range()}), see Section
\ref{sec:PP-on-the}. 

The \texttt{\small{}increments} operator accepts a collection of \emph{n
}number and returns the differences between adjacent numbers. It has
one fewer output elements than input elements. If you feed in a single
element, you get back an Empty piddle (size 0) and if you feed in
an Empty piddle you get back another Empty piddle. There's both a
\texttt{\small{}Code} and a \texttt{\small{}BadCode} segment to demonstrate
BAD handling as well, but the important elements are (a) the \texttt{\small{}RedoDimsCode}
that sets the size of the \texttt{\small{}m} dimension, and (b) the
explicit loop over the m dimension (which could also have been handled
with an explicit C loop and a local index variable). The \texttt{\small{}RedoDimsCode}
uses the \texttt{\small{}\$PDL(}\emph{var}\texttt{\small{})} macro
to get to the underlying C struct. The struct field \texttt{\small{}ndims}
reports the number of dimensions in the PDL; the \texttt{\small{}dims}
array contains the sizes.

(Note that, because the inputs to \texttt{\small{}increments} have
different dimensionality, it cannot handle inplace processing. If
the \texttt{\small{}inplace} flag for in is set, \texttt{\small{}increments}
will ignore it.)
\begin{lyxcode}
{\small{}pp\_def('increments',}{\small \par}
\begin{lyxcode}
{\small{}Pars=>'in(n);~{[}o{]}out(m);',}{\small \par}

{\small{}RedoDimsCode=><\textcompwordmark{}<'EORDC',}{\small \par}
\end{lyxcode}
{\small{}\$PDL\_COMMENT(``~Check~the~dimension~of~'in',~and~calculate~output~size.~``);}{\small \par}

{\small{}if~(~\$PDL(in)->ndims~\&\&~\$PDL(in)->dims{[}0{]}>0~)}{\small \par}
\begin{lyxcode}
{\small{}\$SIZE(m)~=~(~\$PDL(in)->dims(0)-1~)~}{\small \par}
\end{lyxcode}
{\small{}else}{\small \par}
\begin{lyxcode}
{\small{}\$SIZE(m)~=~0;}{\small \par}
\end{lyxcode}
EORDC
\begin{lyxcode}
{\small{}Code=><\textcompwordmark{}<'EOC',}{\small \par}
\end{lyxcode}
{\small{}loop(m)~\%\{}{\small \par}

{\small{}~\$out()~=~\$in(n=>m+1)~-~\$in(n=>m);}{\small \par}

{\small{}\%\}}{\small \par}

{\small{}EOC}{\small \par}
\begin{lyxcode}
{\small{}BadCode=><\textcompwordmark{}<'EOBC',}{\small \par}
\end{lyxcode}
{\small{}loop(m)~\%\{}{\small \par}
\begin{lyxcode}
{\small{}if(\$ISGOOD(n=>m)~\&\&~\$ISGOOD(n=>i))~\{}{\small \par}
\begin{lyxcode}
{\small{}\$out()~=~\$in(n=>m+1)-\$in(n=>m);}{\small \par}
\end{lyxcode}
{\small{}\}~else~\{}{\small \par}
\begin{lyxcode}
{\small{}\$SETBAD(out());}{\small \par}
\end{lyxcode}
{\small{}\}}{\small \par}
\end{lyxcode}
{\small{}\%\}}{\small \par}

{\small{}EOBC}{\small \par}
\begin{lyxcode}
{\small{});}{\small \par}
\end{lyxcode}
\end{lyxcode}
The \texttt{\small{}RedoDimsCode} in \texttt{\small{}increments} explicitly
shortens the output dimension by 1 compared to the input dimension;
the rest just calculates the difference between adjacent elements
and handles BAD values.

\section{\label{sec:Using-PP-With}Using PP With Dataflow}

PP does not only allow you to do explicit, immediate calculations
- your PP operators can set up dataflow links between the input and
output. This is implemented through a quasi-object implemented in
C, called a \emph{trans}. A trans contains all the information needed
to link two PDLs. 

The dataflow engine was implemented by Tuomas Lukka in the early 2000s,
with the intent that you could construct complete behind-the-scenes
data pipelines: modify a particular input PDL, and the result could
automagically cascade through an entire calculation to change the
output PDL. As it turns out, the most common use case for dataflow
has been \emph{multiple representations }of the same underlying data.
Therefore, all of the built-in selection operators in the language
implement dataflow, while the calculation operators do not.

With the information in this section, you can implement dataflow operations
for either selection operators or calculation operators. Dataflow
operators in general construct trans structures that relate their
input and output PDLs, but their PP declarations are nearly the same
as normal computed PP operator declarations. Some additional fields
to \texttt{\small{}pp\_def} are required, to set up the additional
code structures for the underlying trans object.

\subsection{Code/BackCode and a basic dataflow operator}

You can implement a calculated dataflow operator with lazy evaluation,
simply by marking it as a dataflow operator and writing a BackCode
section. Here's an example of a simple two-way computed dataflow operator,
to convert Fahrenheit to Celsius and vice versa.
\begin{lyxcode}
{\small{}pp\_def('FtoC',}{\small \par}
\begin{lyxcode}
{\small{}DefaultFlow~=>~1,~~~\#~Mark~as~a~dataflow~operator}{\small \par}

{\small{}NoPdlThread~=>~1,~~~\#~Turn~off~threadsafe~generated~code~(doesn't~work)}{\small \par}

{\small{}P2Child~=>~1,~~~~~~~\#~Declares~\$PARENT~and~\$CHILD~parameters}{\small \par}

{\small{}Reversible~=>~1,~~~~\#~Enable~BackCode}{\small \par}

{\small{}RedoDims~=><\textcompwordmark{}<'EORD'~\#~No~RedoDims~by~default~-~this~just~copies~\$PARENT~dims~to~\$CHILD.}{\small \par}
\begin{lyxcode}
{\small{}long~ii;}{\small \par}

{\small{}\$SETNDIMS(\$PARENT(ndims));}{\small \par}

{\small{}for(ii=0;~ii<\$PARENT(ndims);~ii++)~\{}{\small \par}
\begin{lyxcode}
{\small{}\$CHILD(dims{[}ii{]})~=~\$PARENT(dims{[}ii{]});}{\small \par}

{\small{}\$CHILD(dimincs{[}ii{]})~=~\$PARENT(dimincs{[}ii{]});}{\small \par}
\end{lyxcode}
{\small{}\}}{\small \par}

{\small{}\$CHILD(datatype)~=~\$PARENT(datatype);}{\small \par}

{\small{}\$SETDIMS();}{\small \par}
\end{lyxcode}
\end{lyxcode}
{\small{}EORD}{\small \par}
\begin{lyxcode}
{\small{}Code~=>~'\$CHILD()~=~(\$PARENT()~-~32)~{*}~5/9;~',}{\small \par}

{\small{}BackCode~=>~'\$PARENT()~=~(\$CHILD()~{*}~9/5)~+~32;~'}{\small \par}
\end{lyxcode}
{\small{});}{\small \par}
\end{lyxcode}
Most of the characters in the operator are in the \texttt{\small{}RedoDims}
section, which copies the parent dimensionality (and internal \texttt{\small{}dimincs}
fields) into the child. This could/should be done by default, but
for PDL versions through 2.009 this particular use case does not work
properly in the PP compiler, so an explicit \texttt{\small{}RedoDims}
is required. The \texttt{\small{}RedoDims} code sets the child's dimensionality
and \texttt{\small{}dimincs} counters to be equal to the parent's.
The \texttt{\small{}Code} converts a single element of the \texttt{\small{}\$PARENT}
from Fahrenheit to Celsius, and stores it in the \texttt{\small{}\$CHILD}.
The BackCode converts a single element of the \texttt{\small{}\$CHILD}
from Celsius to Fahrenheit and stores it in the \texttt{\small{}\$PARENT}.
You use FtoC as follows:
\begin{lyxcode}
{\small{}perldl>~\$F~=~pdl(32);}{\small \par}

{\small{}perldl>~\$C~=~\$F->FtoC;}{\small \par}

{\small{}perldl>~p~\$C;}{\small \par}

{\small{}0}{\small \par}

{\small{}perldl>~\$C++;~p~``\$F~F~=~\$C~C\textbackslash{}n'';}{\small \par}

{\small{}33.8~F~=~1~C}{\small \par}

{\small{}perldl>~\$F++;~p~``\$F~F~=~\$C~C\textbackslash{}n'';}{\small \par}

{\small{}34.8~F~=~1.55555555555555~C}{\small \par}

{\small{}perldl>}{\small \par}
\end{lyxcode}
Assigning to either \texttt{\small{}\$F} or \texttt{\small{}\$C} with
computed assignment (``.=''), or modifying \texttt{\small{}\$F}
or \texttt{\small{}\$C} with side-effect operators, automatically
updates the other one. A dataflow connection has been established.

\subsection{Structure of a dataflow operator}

Dataflow operators make use of several stages of code execution. In
a typical use case, at call time the operator defines a trans that
involves computing some input parameters and/or examining input data.
Then it sets the dimensions of the output PDL. The actual calculation
is deferred until the output PDL is used. This ``lazy evaluation''
is a wart that's left over from the development history of dataflow
in PDL, but could in principle be extended to elementwise lazy evaluation.
(At present, all operators calculate the results for every element
in a PDL, as soon as the first element is accessed). These different
steps are implemented with different \texttt{\small{}pp\_def} fields. 

Dataflow operations work by creating a separate structure/object called
a ``\texttt{\small{}trans}'' that relates two PDLs -- the parent
and the child. The trans structure includes metadata about the transformation
and also pointers to the code blocks you declare in the \texttt{\small{}pp\_def}
call. Its elements are accessible with the \texttt{\small{}\$PRIV()}
macro, which expands to C code that references the private C struct.

The \texttt{\small{}pp\_def} fields you need to know for a basic dataflow
operator are:
\begin{description}
\item [{\texttt{\small{}Pars}}] The signature (as in a regular computed
operator). There are some caveats -- see ``\texttt{\small{}P2Child}''
and ``\texttt{\small{}DefaultFlow}'', below.
\item [{\texttt{\small{}OtherPars}}] Any additional parameters not subject
to the usual threading rules (which could be the main parameters --
see the \texttt{\small{}rangeb} operators in \texttt{\small{}slices.pd}
for an example)
\item [{\texttt{\small{}Doc}}] POD format documentation for the operator
\item [{\texttt{\small{}HandleBad}}] Same as for a regular computed operator
-- ignore for default treatment, set to 1 to enable the \texttt{\small{}BadCode}
(and \texttt{\small{}BadBackCode)} fields, set to 0 to cause the code
to barf if it receives a PDL with the badflag set.
\item [{\texttt{\small{}Code}}] Just like a normal PP operator, this computes
the output PDL value from the input PDL values. 
\item [{\texttt{\small{}BackCode}}] This should compute (and set) the value
of the input PDL from the output PDL. It gets called if the output
PDL is changed by some other operator, to flow data back into the
input PDL. You do not need to specify BackCode for one-way flow. (See
the \texttt{\small{}Reversible} flag, below)
\item [{\texttt{\small{}DefaultFlow}}] You almost certainly want to set
this flag to 1. It sets up a dataflow relationship between two special
parameters, ``\texttt{\small{}PARENT}'' and ``\texttt{\small{}CHILD}'',
that should be declared in the \texttt{\small{}Pars} field (or in
the \texttt{\small{}P2Child} field, see below). If you don't either
use the \texttt{\small{}P2Child} flag or declare \texttt{\small{}PARENT}
and \texttt{\small{}CHILD} in the Pars field, PP will throw an error.
\item [{\texttt{\small{}Reversible}}] This flag indicates whether the transformation
is two-way. If it is, then you need to specify both \texttt{\small{}Code}
and \texttt{\small{}BackCode} segments (or \texttt{\small{}EquivCPOffsCode}
-- see below).
\item [{\texttt{\small{}P2Child}}] This flag takes the place of \texttt{\small{}Pars}
for simple dataflow. Setting the flag is equivalent to declaring\texttt{\small{}
Pars=>'PARENT(); {[}oca{]}CHILD()};'. It declares a zero-dimensional
input \texttt{\small{}PARENT} and a zero-dimensional output \texttt{\small{}CHILD}. 
\item [{\texttt{\small{}NoPdlThread}}] This is a flag that prevents the
normal threading behavior of PP. If it is set, no threadloop will
be executed around (or within) the various \texttt{\small{}Code} blocks
where it would normally happen. That is useful for edge cases where
you might want to work independently of the normal threading mechanism
(see Chapter \ref{sec:PP-on-the}).
\item [{\texttt{\small{}Comp}}] Should be a C declaration of some fields
to be computed (and stashed) during setup of a trans. If this is present,
then the \texttt{\small{}OtherPars} parameters are not automatically
incorporated in the array as with conventional computed operators.
(That happens with a default \texttt{\small{}Comp} and \texttt{\small{}MakeComp}
field that gets overridden by your \texttt{\small{}Comp} and \texttt{\small{}MakeComp}.)
See \texttt{\small{}MakeComp} for details. The \texttt{\small{}Comp}
field has (limited) access to the \texttt{\small{}\$COMP(}) macro
itself, specifically so that you can declare variable-length arrays
here. In particular, constructs like ``\texttt{\small{}int b; int
a{[}\$COMP(b){]}}'' work. Before using those arrays (in the corresponding
\texttt{\small{}MakeComp} code block) you have to set the variables
that hold the sizes, then run the macro \texttt{\small{}\$DOCOMPDIMS()}.
That allocates all the variable length arrays.
\item [{\texttt{\small{}MakeComp}}] Should contain code to compute and
stash the contents of the \texttt{\small{}\$COMP()} macro, called
the computed stash. This is the place to store any state or parametric
information about the trans itself. The \texttt{\small{}OtherPars}
parameters are accessible in this block of code as regular C identifiers
-- so if you pass in an OtherPar called \texttt{\small{}foo}, you
access it here as \texttt{\small{}foo} and not as \texttt{\small{}\$COMP(foo)}.
If you don't specify \texttt{\small{}MakeComp}, you get a default
\texttt{\small{}MakeComp} block that just copies any \texttt{\small{}OtherPars}
into the \texttt{\small{}\$COMP()} structure.
\item [{\texttt{\small{}RedoDims}}] Should contain code to compute and
stash the dimensions of the child PDL, from the dimensions of the
parent PDL and the parameters passed in.
\item [{\texttt{\small{}EquivCPOffsCode}}] This field is useful for simple
dataflow relating arbitrary locations in one PDL to arbitrary locations
in another, for example inside the built-in \texttt{\small{}range}
and \texttt{\small{}index} operators. It contains code for relating
particular elements between the two linked PDLs (parent and child).
Its name derives from ``equivalent child/parent offsets'', and the
code should implement direct-copy dataflow between a single element
of the special \texttt{\small{}\$CHILD()} and \texttt{\small{}\$PARENT()}
piddles. If present, \texttt{\small{}EquivCPOffsCode} replaces \texttt{\small{}Code},
\texttt{\small{}BackCode}, and (if \texttt{\small{}HandleBad} is set)
\texttt{\small{}BadCode} and \texttt{\small{}BadBackCode} -- these
all end up running the \texttt{\small{}EquivCPOffsCode}, with the
appropriate assignment placed inside to slosh data forward or backward.
The \texttt{\small{}EquivCPOffsCode} should loop over all values for
which dataflow should happen. For each of those values, it needs call
one of the special macros \texttt{\small{}\$EquivCPOffs(}\emph{pi,ci}\texttt{\small{})}
or \texttt{\small{}\$EquivCPTrunc(}\emph{pi,ci,oob}\texttt{\small{})}
for each value. Here, \emph{pi} is the computed offset in the parent
to a particular element, \emph{ci} is the offset to the corresponding
element in the child, and \emph{oob} is a flag indicating that the
offset is out of bounds in the parent (and should therefore be loaded
with BAD or 0 in the child for forward flow, or ignored for reverse
flow). Note that \texttt{\small{}EquivCPOffsCode}, unlike \texttt{\small{}Code}
and \texttt{\small{}BackCode}, does not deal with threading! You have
to do the looping yourself.
\end{description}
The calling structure is as follows. When you first invoke the operator,
\texttt{\small{}MakeComp} gets called (making the \texttt{\small{}\$COMP()}
variable macros available), then \texttt{\small{}RedoDims}. Only when
the user makes use of the resulting child PDL does the \texttt{\small{}EquivCPOffsCode}
(or \texttt{\small{}Code} or \texttt{\small{}BackCode}) get called. 

Good examples of dataflow operators are to be found in the PDL distribution
itself, in ``...\texttt{\small{}/Basic/Slices/slices.pd}''. Check
out the code for \texttt{\small{}slice} and \texttt{\small{}sliceb},
or \texttt{\small{}range} and \texttt{\small{}rangeb}. Those operators
are implemented without the \texttt{\small{}PMCode} field, because
\texttt{\small{}PMCode} seems to not work correctly for dataflow operators.
The \texttt{\small{}sliceb} operator uses a special kind of trans
called an ``affine'' trans, which doesn't actually use calculation
code or \texttt{\small{}EquivCPOffs} at all -- it causes the child
PDL to point directly at the parent's data structure. The \texttt{\small{}rangeb}
operator generates a data structure containing information about each
range to be extracted, then walks through the source data calling
\texttt{\small{}EquivCPOffs}.

\section{\label{sec:PP-on-the}PP on the Edge (things PP can't do easily,
and how to do them)}

PP is great for performing ordinary vectorized calculations where
the dimensions are known or computable up front. There are certain
problem domains at which it is awkward -- for example, allowing an
operator to vary its own signature at runtime (changing the number
of active dims) is difficult, as is accepting a variable length argument
list. Here are some basic strategies for writing operators that are
right at the edge of what PP can do easily.

\subsection{Variable Active Dims}

The most obvious example of an operator with variable active dims
is the built-in \texttt{\small{}indexND}. That operator collapses
an index variable by lookup into a source array: the 0 dim of the
index is treated as running across dimension in the source array,
and each row of the index variable causes one element of the source
array to be indexed. Depending on the size of the index variable's
active dim, the signature should look like one of the following:
\begin{lyxcode}
{\small{}PARENT(i0);~~~~~~~~~~index(n=1);~{[}o{]}CHILD());}{\small \par}

{\small{}PARENT(i0,i1);~~~~~~~index(n=2);~{[}o{]}CHILD());}{\small \par}

{\small{}PARENT(i0,i1,i2);~~~~index(n=3);~{[}o{]}CHILD());}{\small \par}

{\small{}PARENT(i0,i1,i2,i3);~index(n=4);~{[}o{]}CHILD());}{\small \par}

{\small{}\#~etc.}{\small \par}
\end{lyxcode}
PP can't handle that type of operator easily and straightforwardly,
because the shape of \texttt{\small{}PARENT} depends on the size of
one of the dimensions in \texttt{\small{}index}.

The meat behind \texttt{\small{}indexND} is the slightly more general
\texttt{\small{}rangeb} operator, which pulls a specified rectangular
block out of the source PDL at each indexed location. It can be found
in the PDL source distribution in the file ``\texttt{\small{}Basic/Slices/slices.pd}''.
It works by sidestepping the threading behavior entirely. The \texttt{\small{}PARENT}
and \texttt{\small{}CHILD} are passed in with 0 active dims via the
\texttt{\small{}P2Child} flag to \texttt{\small{}pp}\_def (which autodeclares
a basic parent/child signature). All other parameters (in particular
the index and size) are declared as \texttt{\small{}OtherPars} with
type ``\texttt{\small{}SV {*}}'' so that the code has direct access
to the Perl variable containing the index. PDL provides a function\texttt{\small{},
PDL->SvPDLV}(), that accepts a \texttt{\small{}SV {*}} and returns
the corresponding \texttt{\small{}pdl {*}}, so you can access the
internal structure directly. 

The \texttt{\small{}rangeb} operator has three key parts: the \texttt{\small{}MakeComp}
code gets executed first, and extracts all the relevant size and offset
information from the input PDLs into C arrays of known size. Then
the \texttt{\small{}RedoDims} code block calculates the size, dimincs
(memory stride for each dimension) and datatype of the \texttt{\small{}CHILD}
(output) variable, from the input. It uses the \$\texttt{\small{}SETDIMS}()
macro to allocate the child's actual dimensions and data block, and
reproduces the regular threading rules used by PP. Finally, the \texttt{\small{}EquivCPOffsCode}
gets called to copy values into the new \texttt{\small{}CHILD} piddle.
The \texttt{\small{}EquivCPOffsCode} for \texttt{\small{}rangeb} iterates
explicitly over all input dimensions including thread dimensions.
Iterating over a variable number of dimensions is tricky; there's
a nice example at the bottom of the EquivCPOffsCode for rangeb. The
main loop is a \texttt{\small{}do/while} construct, with an explicit
iterator at the end. The iterator is a \texttt{\small{}for} loop that
handles carry from the fastest-moving to the next-fastest-moving index,
and on up to the slowest-moving index.

\subsection{Variable Argument Counts}

The best way to handle a variable argument count is to use a Perl-side
currying function (\texttt{\small{}PMCode} in a module or a separately
declared function in an inline script) that stuffs the variables into
a Perl array/list, then pass a ref to the array into the PP operator
as an \texttt{\small{}OtherPars} parameter. Your code can dereference
the ref and traverse the array (\texttt{\small{}AV {*}}) directly.
Each element of the AV will be a separate \texttt{\small{}SV {*}}
that you can parse with your own code, and/or convert to a\texttt{\small{}
pdl {*}} . See the \texttt{\small{}perlguts} and/or \texttt{\small{}perlapi}
man page for details on how to manipulate Perl AVs.

\subsection{Returning Perl Variables}

There is no explicit SV return mechanism in the \texttt{\small{}Code}
blocks of a PP function If you need to return a single SV value to
a Perl function, you can pre-allocate a variable with a Perl-side
currying function (\texttt{\small{}PMCode} in a module or a separately
declared function in an inline script), then pass in a ref to the
value as an \texttt{\small{}OtherPars} parameter. Your PP code can
dereference the ref to get an lvalue SV, and assign to that SV. Then
the currying function can return the variable's value.

The same mechanism works for lists -- if you need to return several
SVs, you can construct an array in your currying function and pass
a ref to it into the actual PP code. Then the C side can populate
the referenced AV with SVs to be returned by the currying function.

\section{PDL internals: accessing PDLs directly from C}

PDL data structures are directly accessible from your C code within
PerlXS. Most times you want to use PP to access the structure, but
(particularly if you are using an external library) you may want to
get in and mess around with the internals yourself. Here is a brief
overview to get you going. The most powerful way to access PDLs is
via PP (Section \ref{sec:Basic-PP}), but both direct access and a
module called \texttt{\small{}CallExt} offer slightly lighter-weight
forms of access.

From within Perl, PDLs are blessed scalar refs. The scalar that is
referenced contains an IV (Perl integer value) that is itself a pointer
to a C structure (\texttt{\small{}struct pdl}). If you \#include both
\texttt{\small{}<pdl.h>} and \texttt{\small{}<pdlcore.h>}, you get
access to the data structures and to a variety of important access
routines. You do \emph{not} need to link to a PDL library to access
this rather extensive C API. 

When Perl executes \texttt{\small{}``use PDL;}'', a ``core'' structure
(called ``\texttt{\small{}PDL}'' on the C side) is created that
contains direct function pointers to the utilities. The core structure
is a static patchtable that allows all interested modules to call
the same C functions -- otherwise the Perl dynamic linking code would
link a separate copy of the PDL core functions for each module that
uses them. If you are using PP, a pointer to this patchtable is automatically
placed into a C file-scope static variable called ``\texttt{\small{}PDL}''
-- hence utility routines are accessed from C as (e.g.) ``\texttt{\small{}PDL->SvPDLV(foo)}''. 

{[}If you want to use the API from a non-PP XS module, then the header
section of your PerlXS file should declare a \texttt{\small{}struct
pdlcore {*}} called \texttt{\small{}PDL}, and the \texttt{\small{}BOOT:}
section of your perlXS file should retrieve the value of this pointer
from the Perl global variable \texttt{\small{}\$PDL::SHARE}. (You
must also ensure that the pointer has been initialized, e.g. with
a ``\texttt{\small{}use PDL;}'' in the Perl portion of your module).{]}

For a full list of the routines in the PDL C API, you currently need
to examine the online documentation and source code comments. A separate
document is \emph{planned} to describe these PDL internals, but the
online documentation is extensive and should get you going.

The very simplest, and preferred, way to access the data in a PDL
from C/XS is to wrap your code in a PP call and use the PP macros.
For cases where that will not work easily, you can have your routine
accept an SV {*} (Perl scalar) and run \texttt{\small{}PDL->SvPDLV}
on the \texttt{\small{}SV {*}}. The return value is a \texttt{\small{}pdl
{*}}, i.e. a pointer to the actual PDL internal structure. Before
accessing the data directly from C without the PP macros, you must
check the status flags, non-nullness of the data field, and data type
of the PDL. You will also need to cast the \texttt{\small{}data} field
to the appropriate type of pointer before dereferencing it. The data
are structured using the \texttt{\small{}dimincs} array of the PDL
structure -- you multiply each element of dimincs by the index you
want in the corresponding dimension, and summing those products gives
you the offset into \texttt{\small{}data}.

\subsection{\label{sub:The-struct-pdl}The struct pdl type}

The typedefed ``\texttt{\small{}pdl}'' type is a \texttt{\small{}``struct
pdl}'', which is declared in \texttt{\small{}pdl.h} in the PDL distribution.
Here is a list of the fields in it. Nonstandard types are declared
in \texttt{\small{}pdl.h}. The \texttt{\small{}PDL\_Indx} type, in
particular, is either a 32 bit or 64 bit unsigned value depending
on your sysem architecture. 
\begin{description}
\item [{magicno}] (unsigned long) An unsigned long with a magic number
(\texttt{\small{}PDL\_MAGICNO}; declared in \texttt{\small{}pdl.h})
in it. This is a fence for debugging purposes.
\item [{state}] (int) A bit field containing flags that describe the state
of the PDL. The flags are declared in \texttt{\small{}pdl.h}: use
bitwise-and to mask them out of the state.

\begin{description}
\item [{\texttt{\small{}PDL\_ALLOCATED}}] Data have been allocated for
this PDL. Also implies that the data is up-to-date (unless the PDL\_PARENTDATACHANGED
flag is set).
\item [{\texttt{\small{}PDL\_PARENTDATACHANGED}}] This PDL is connected
to another one via dataflow, and that other PDL has changed -- but
this one has not yet been updated.
\item [{\texttt{\small{}PDL\_PARENTDIMSCHANGED}}] This PDL is connected
to another one via dataflow, and that other PDL's shape has changed
-- but this one has not yet been updated.
\item [{\texttt{\small{}PDL\_PARENTREPRCHANGED}}] The representation of
the parent changed (for example, \texttt{\small{}sever()} broke it
from its parent), so data access hooks like the \texttt{\small{}incs}
field need to be recalculated.
\item [{\texttt{\small{}PDL\_ANYCHANGED}}] This is the bitwise OR of the
previous 3 flags - i.e. the parent has been touched.
\item [{\texttt{\small{}PDL\_DATAFLOW\_F}}] Track forward dataflow starting
from this PDL into its children.
\item [{\texttt{\small{}PDL\_DATAFLOW\_B}}] Track reverse dataflow back
from this PDL into its parent.
\item [{\texttt{\small{}PDL\_DATAFLOW\_ANY}}] bitwise OR of the previous
2 flags - i.e. this PDL has a dataflow connection somewhere.
\item [{\texttt{\small{}PDL\_NOMYDIMS}}] this PDL is null (auto-reshaping
to match needed shape in an expression).
\item [{\texttt{\small{}PDL\_MYDIMS\_TRANS}}] the dims are received through
a trans (transformation structure) from another PDL.
\item [{\texttt{\small{}PDL\_OPT\_VAFFTRANSOK}}] It's okay to attach a
virtual-affine trans to this PDL (i.e. to point another PDL at the
same data block)
\item [{\texttt{\small{}PDL\_HDRCPY}}] The hdrcpy flag (causes the header
of this PDL to be autocopied into result PDLs when appropriate, in
expressions and operators)
\item [{\texttt{\small{}PDL\_BADVAL}}] The BAD flag for this PDL - if set,
the PDL may contain BAD values and the BadCode gets executed instead
of the normal Code.
\item [{\texttt{\small{}PDL\_TRACEDEBUG}}] causes a bunch of PP internal
debugging
\item [{\texttt{\small{}PDL\_INPLACE}}] the inplace flag for this PDL -
if set, the PDL has been primed for an in-place operation by the calling
entity. 
\item [{\texttt{\small{}PDL\_DESTROYING}}] is set to indicate that this
PDL is in the process of being destroyed, so data flow operations
should be ignored.
\item [{\texttt{\small{}PDL\_DONTTOUCHDATA}}] indicates that the \texttt{\small{}data}
pointer is inviolate - don't change it, free it, nor use the datasv.
Used mainly for piddles that mmap memory from files.
\end{description}
\item [{trans}] (pdl\_trans {*}) This is a pointer to a transformation
structure, which implements dataflow for PDLs that require explicit
flow. The transformation structure includes code refs that are called
by PP whenever this PDL is updated. It is NULL if the PDL has no dataflow
``children''. Complicated dataflow (such as indexing, dicing, and
ranging) uses this mechanism. 
\item [{vafftrans}] (pdl\_vaffine {*}) This is a pointer to a more optimized
transformation structure for PDLs with dataflow ``children'' that
are related by an affine transformation. Affine transformations are
transformations in which the indices in the child are related linearly
to the corresponding indices in the parent (as in simple slicing).
They use a different structure and are very fast, because no copying
is needed -- the same physical array in memory can be indexed by the
related PDLs.
\item [{sv}] (void {*}) If this is non-NULL, it is a pointer back to the
Perl SV that contains the PDL. 
\item [{datasv}] (void {*}) This is either NULL or a pointer to a string
Perl SV that contains the data in its data field. PDLs use the Perl
memory management infrastructure, so the data block of a given pdl
{*} is stored as a byte string in an anonymous Perl SV somewhere.
The datasv should have a reference count of 1 for each PDL that accesses
it. The ref count gets decremented when the PDL is destroyed, freeing
the memory.
\item [{data}] (void {*}) This is the actual data pointer, and points to
the string field in the associated SV. It can be null if the PDL contains
no actual data. The type of the data depends on the value of the \texttt{\small{}datatype}
field, below.
\item [{badvalue}] (double) If this PDL uses BAD value logic, then this
is the value that is considered BAD. It is stored as a double regardless
of data type of the PDL itself.
\item [{has\_badvalue}] (int) This is a simple flag -- 0 for straight-up
values, 1 for PDLs that recognize bad values.
\item [{nvals}] (PDL\_Indx) This is the total number of elements contained
in the whole PDL -- it's the product of the elements of the ``dims''
array. It is returned to Perl by \texttt{\small{}PDL::nelem}. 
\item [{datatype}] (int) indicates the data type of the PDL. It's declared
as an int, but is actually an enum - access it with one of the datatype
enums in pdl.h (\texttt{\small{}PDL\_B}, \texttt{\small{}PDL\_S},
\texttt{\small{}PDL\_US}, \texttt{\small{}PDL\_L}, \texttt{\small{}PDL\_IND},
\texttt{\small{}PDL\_LL}, \texttt{\small{}PDL\_F}, or \texttt{\small{}PDL\_D}). 
\item [{dims}] (PDL\_Indx {*}) is a pointer to an array of sizes for each
dimension of the data -- the contents are returned to Perl by \texttt{\small{}PDL::dims}
or \texttt{\small{}PDL::shape}. This can be allocated separately or
point to a small preallocated space farther down, depending on the
number of dimensions.
\item [{dimincs}] (PDL\_Indx {*}) is a pointer to an array that caches
the memory increments associated with each dimension in the PDL. Like
\texttt{\small{}dims}, it can be separately allocated or simply a
part of this \texttt{\small{}struct pdl}, depending on how many dimensions
there are.
\item [{ndims}] (short) is the length of the dim list for this PDL. It
is returned to Perl by \texttt{\small{}PDL::ndims}.\end{description}

\end{document}